\documentclass[aps, pra, amsmath, amssymb, superscriptaddress, nofootinbib]{revtex4-2}
\usepackage[english, brazilian]{babel}
\usepackage[utf8]{inputenc} 
\usepackage[T1]{fontenc}
\usepackage{graphicx}
\usepackage{subcaption}
\usepackage{dcolumn}
\usepackage{bm}
\usepackage[bottom]{footmisc}
\usepackage{blindtext}
\usepackage{float}
\usepackage{xcolor}

\usepackage[version=4]{mhchem}

\begin{document}

\title{Determinação da Distância à Grande Nuvem de Magalhães Através das Estrelas Variáveis Cefeidas Disponíveis no Catálogo OGLE-IV}

\author{Kevin Mota da Costa}
\email{kevin.costa@edu.ufes.br}
\affiliation{Departamento de Física, Universidade Federal do Espírito Santo, 29075-910 Vitória - ES, Brasil}

\author{Alan Miguel Velasquez-Toribio}
\email{alan.toribio@ufes.br}
\affiliation{Departamento de Física, Universidade Federal do Espírito Santo, 29075-910 Vitória - ES, Brasil}

\author{Júlio César Fabris}
\email{julio.fabris@ufes.br}
\affiliation{Departamento de Física, Universidade Federal do Espírito Santo, 29075-910 Vitória - ES, Brasil}
\affiliation{National Research Nuclear University MEPhI, Kashirskoe sh. 31, Moscow 115409, Russia}

\date{\today}

\begin{abstract}
\begin{center}
    \textbf{Resumo}
\end{center}
    
Neste trabalho, é discutida a determinação da distância até a Grande Nuvem de Magalhães (GNM) através da Lei de Leavitt, utilizando o catálogo público de estrelas Variáveis Cefeidas Clássicas do projeto observacional OGLE-IV (The Optical Gravitational Lensing Experiment Collection of Variable Stars), que consiste em 4709 estrelas na Grande Nuvem de Magalhães. Para determinar o período de pulsação das estrelas Variáveis Cefeidas, utilizamos o algoritmo computacional "Periodograma Lomb-Scargle" modificado para nossos dados. Adicionalmente, com o cálculo do período, podemos derivar uma relação período-luminosidade para as Variáveis Cefeidas da Grande Nuvem de Magalhães e, com o uso de uma distância de calibração independente, derivar seus módulos de distância. Também discutimos alguns conceitos teóricos gerais do mecanismo físico de oscilação das estrelas variáveis.

\textbf{Palavra-chave}: relação período-luminosidade, estrelas Variáveis Cefeidas, Grande Nuvem de Magalhães.

\selectlanguage{english}

\begin{center}
    \textbf{Abstract}
\end{center}
    
In this work, we discuss the determination of the distance to the Large Magellanic Cloud (LMC) using the Leavitt Law, utilizing the public catalog of Classical Cepheid Variable stars from the observational project OGLE-IV (The Optical Gravitational Lensing Experiment Collection of Variable Stars), consisting of 4709 stars in the Large Magellanic Cloud. To determine the pulsation period of Cepheid Variable stars, we employ the computational algorithm \textit{Lomb-Scargle periodogram} modified for our data. Additionally, with the calculation of the period, we can derive a period-luminosity relation for Cepheid Variables in the Large Magellanic Cloud and, using an independent calibration distance, deduce their distance moduli. We also discuss some general theoretical concepts of the physical mechanism behind the oscillation of variable stars.


\selectlanguage{brazilian}

\end{abstract}

\maketitle 

\section{Introdução}

Com o início das investigações sistemáticas dos fenômenos físicos através do método científico, introduzido por Galileu(1564 - 1642) \cite{hodgson2003}, a ciência passou a exigir critérios rigorosos para progredir, focando especialmente na verificação experimental das previsões teóricas, o que é fundamental para balizar nosso conhecimento
sobre a natureza.

\par Consequentemente é evidenciada a necessidade de ferramentas que possibilitam a observação dos fenômenos físicos. Contudo, surge a questão: como abordar fenômenos que nos fogem aos sentidos? Por exemplo, como é possível medir distâncias até galáxias tendo em vista que não podemos fazer nenhuma medida direta? Essa é uma das questões que esse trabalho pretende discutir.

\par Nossa jornada começa em 3 de agosto de 1596, quando o astrônomo David Fabricius (1564-1617) observou uma certa estrela para determinar a posição de um planeta do sistema solar. Sua observação continuou até 21 de agosto, quando a magnitude aparente\footnote[1]{A magnitude aparente representa a luminosidade de um objeto celeste quando observado da Terra, enquanto a magnitude absoluta denota o brilho intrínseco da estrela. Quanto menor a magnitude absoluta, maior é o brilho da estrela, indicando uma relação inversamente proporcional entre a magnitude absoluta e a luminosidade intrínseca. Ver Seção \ref{sec:determinando.distancia.grande.nuvem.magalhaes.via.lactea}.} da estrela se passou de 3 para 2; em setembro a estrela desapareceu completamente do céu. Fabricius pensou que tinha observado uma NOVA, até que em 15 de fevereiro de 1609 a estrela reapareceu no céu \cite{hoffleit1997}. Essa foi a primeira detecção das denominadas "estrelas variáveis".

\par O estudo das estrelas variáveis, apesar de intrigante, não proporcionou grandes revoluções na astronomia, cosmologia e astrofísica, até que em 1907 a astrônoma Henrietta Leavitt, ao catalogar 1777 estrelas Variáveis Cefeidas na Pequena Nuvem de Magalhães, verificou experimentalmente que existe uma correlação entre o período de pulsação dessas estrelas com a sua magnitude aparente (denominada relação período-luminosidade, ou Lei de Leavitt ) \cite{leavitt1908, machado2021estrelascefeidas}. Essa descoberta foi fundamental para o surgimento da cosmologia moderna. A Lei de Leavitt pode ser representada pela equação \begin{equation}
    m = a \log \left({P}\right) + b,
    \label{eq:leavitt}
\end{equation} onde $m$ é a magnitude aparente da estrela e $P$ é o período de pulsação da estrela em Dia Juliano\footnote[2]{O Dia Juliano é uma contagem contínua de dias e frações de dia desde 1º de janeiro de 4713 a.E.C \cite{almanac2023}.}. As constantes $a$ e $b$ são intrínsecas à galáxia hospedeira das estrelas e à faixa espectral na qual a magnitude é observada.

Em 1926, o astrônomo Edwin Hubble utilizou o telescópio de $2,5$ metros localizado em Monte Wilson para determinar a distância entre a galáxia de Andrômeda e a Via Láctea. Ele baseou-se na relação período-luminosidade de Leavitt para realizar essa medição. O resultado de suas observações indicou uma distância aproximada de $700.000$ anos-luz\footnote[3]{Vale ressaltar que o valor atualmente aceito é de aproximadamente $2.5$ milhões de anos-luz \cite{jpl_andromeda_info}.}.  A comparação dessa distância com as estimativas da época para o tamanho da Via Láctea, que eram da ordem de $100.000$ anos-luz\footnote[4]{Conforme citado em \cite{brennan2019milkyway}, estimativas da época indicavam esse valor.}, levou Hubble à conclusão de que Andrômeda não fazia parte da Via Láctea, sendo uma outra galáxia..

A confirmação da existência de novas galáxias pelo trabalho de Hubble marcou um momento crucial na história da astronomia \cite{NYT1926}. Até 1926, havia um acalorado debate sobre a presença de outras galáxias no universo \cite{florio2021via}. A descoberta de Hubble trouxe uma revolução à astronomia moderna, resultando em avanços de suma importância. Isso incluiu a formulação da lei observacional de Hubble-Lemaître, que deu significado às equações de Friedman e, consequentemente, conduziu à criação do que hoje chamamos de Modelo Cosmológico Padrão. Atualmente, o Modelo Cosmológico Padrão é amplamente aceito na comunidade científica como o modelo predominante para explicar o funcionamento do Cosmo.

\par No diagrama de Hertzsprung-Russell, que relaciona a magnitude absoluta das estrelas com sua temperatura superficial, as Variáveis Cefeidas - assim como outras estrelas variáveis - se encontram nas faixas de instabilidade, características de estrelas pulsantes \cite{madore98}.

\begin{figure}[H]
    \centering
    \includegraphics[scale=0.5]{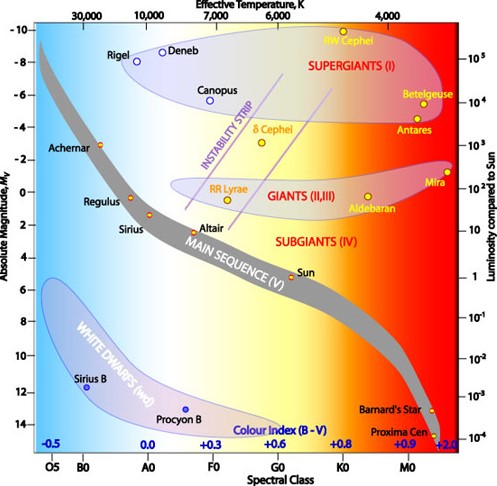}
    \caption{Diagrama de Hertzsprung-Russell. Fonte: R. Hollow, CSIRO. \cite{hr_diagram}}
    \label{fig:diagrama.hr}
\end{figure}

As Variáveis Cefeidas são estrelas supergigantes com luminosidade na faixa de $500-30.000 L_\odot$ \footnote[5]{$L_\odot = 382,8 \cdot 10^{24} J / s $ é a luminosidade solar \cite{nasa-sunfact}.}  e temperatura superficial da ordem da temperatura superficial solar \footnote[6]{O valor atual é da ordem de $5772 K$ \cite{nasa-sunfact}.}. Essas estrelas realizam pulsações radiais com da ordem de $1-50 \, \mathrm{dias}$, e podem ser observadas através de longas distâncias. A periodicidade da curva de luz de uma Variável Cefeida é semelhante à periodicidade de sua curva de velocidade radial, onde a velocidade radial mínima ocorre quando a estrela apresenta magnitude máxima. A magnitude típica das estrelas Variáveis Cefeidas é da ordem de $0.5-2$ no visível, e a velocidade radial é da ordem de $30-60 \mathrm{km} \, \mathrm{s}^{-1}$ \cite{evans-ceph-variable}.

\par Eddington, em 1917, publicou um trabalho argumentando que as estrelas Variáveis Cefeidas são estrelas que realizam pulsação radial devido a suas características termodinâmicas, com o funcionamento análogo ao de uma máquina térmica. Esse mecanismo foi aprimorado por diversos físicos, como J.P. Cox, S.A. Zhevakin, Robert F. Christy, entre outros, proporcionando uma melhor compreensão sobre as estrelas variáveis: a energia é armazenada quando ocorre a produção do Hélio duas vezes ionizado (durante o estágio de compressão do ciclo) e então liberada conforme o Hélio se recombina durante o estágio de expansão. É importante destacar que a pulsação é uma característica do envoltório externo da estrela e não está relacionada à taxa de produção de energia termonuclear no seu núcleo \cite{evans-ceph-variable}.

\par As curvas de luz, que representam a relação entre a luminosidade das estrelas e o tempo, como ilustrado na Figura \ref{fig:curva.luz.fase}, geralmente exibem uma assimetria característica. Elas apresentam um rápido aumento em direção à luminosidade máxima, seguido por um declínio mais gradual. Essa forma da curva de luz varia de forma previsível com o período, seguindo um padrão conhecido como progressão de Hertzsprung \cite{evans-ceph-variable}.

\par Nesse contexto, é interessante notar que surge uma saliência no trecho descendente da curva de luz em estrelas com períodos de aproximadamente uma semana. À medida que os períodos das estrelas se tornam progressivamente mais longos, essa saliência é observada em fases anteriores, aproximando-se da luminosidade máxima em estrelas com períodos de 10 dias, onde podem até mesmo ocorrer dois picos de luminosidade. Entretanto, à medida que os períodos das estrelas se tornam ainda mais longos, essa protuberância se desloca para o trecho ascendente da curva \cite{evans-ceph-variable}.

Há várias teorias para explicar essas saliências. Uma delas sugere que elas podem ser resultado de um eco gerado pelas pulsações na superfície do núcleo das estrelas. Outra explicação alternativa propõe que as saliências são o resultado de uma ressonância, ocorrendo quando o segundo período harmônico da estrela é aproximadamente a metade do período fundamental \cite{evans-ceph-variable}.

\par A faixa evolutiva de massa típica das estrelas Variáveis Cefeidas, excluindo aquelas com períodos extremamente longos, implica massas da ordem de $3-9 M_\odot$ \footnote[7]{$M_\odot = 1,989 \cdot 10^{30} kg$ é a massa solar \cite{nasa-sunfact}.}. Além disso, uma dada Variável Cefeida pode atravessar a faixa de instabilidade no diagrama de Hertzsprung-Russell várias vezes durante a sua evolução. Inicialmente, a estrela deixa a sequência principal quando o hidrogênio (\(H\)) em seu núcleo se esgota. Durante esse estágio, o \(H\) é consumido em uma camada que envolve um núcleo de hélio (\(He\)) temporariamente inerte. Sob a intensa pressão das reações nucleares na camada de hidrogênio ao redor do núcleo de hélio, a estrela experimenta uma expansão. Nessa fase, a estrela continua a queimar hidrogênio em uma casca ao redor do núcleo de hélio, levando à diminuição da temperatura superficial da estrela. Como a temperatura das estrelas está intrinsecamente relacionada com sua cor, neste caso, a estrela esfria e aumenta de tamanho, transformando-se em uma gigante vermelha \cite{knapp2011ast403, evans-ceph-variable}.

    \par Posteriormente, a estrela atravessa rapidamente a faixa de instabilidade, seguindo uma escala de tempo conhecida como escala de Kelvin-Helmholtz, que descreve o tempo necessário para uma estrela colapsar sob a influência da gravidade após o fim do processo de fusão nuclear \cite{knapp2011ast403}. Ela então ascende pelo ramo das gigantes vermelhas. Após a ignição da queima de \ce{He} em seu núcleo, a estrela pode realizar um ciclo de aumento de temperatura no diagrama de Hertzsprung-Russell. Esse ciclo pode alcançar temperaturas suficientemente elevadas para atravessar novamente a faixa de instabilidade, resultando em mais dois cruzamentos subsequentes \cite{evans-ceph-variable}.

\par A queima de hélio no núcleo é um estágio evolutivo relativamente longo, e a estrela pode permanecer na faixa de instabilidade por muito mais tempo (talvez por um fator de $50$) do que no primeiro cruzamento. A localização exata dos ciclos é uma função da massa e da composição química, de modo que acima e abaixo da massa mais favorável, o número de Variáveis Cefeidas diminuirá rapidamente. As estrelas mais massivas evoluem mais rapidamente em qualquer caso, de modo que o tempo de residência de pico corresponderá a uma massa relativamente baixa com um declínio no número de estrelas de período mais longo, o que é acentuado pela relativa raridade de estrelas massivas. Todas as Variáveis Cefeidas nos extremos da distribuição de massa - e, consequentemente, período - devem estar em seu primeiro cruzamento da faixa de instabilidade. Estima-se que essas estrelas representem cerca de 10\% de todas as Cefeidas \cite{evans-ceph-variable}.

\par No presente trabalho, vamos determinar a distância média à Grande Nuvem de Magalhães através das estrelas Variáveis Cefeidas. Na Seção \ref{sec:modelo.pulsacao.estelar}, é discutido um mecanismo simplificado de pulsação de estrelas variáveis, mostrando que é possível modelar uma estrela variável com período de oscilação fixo. Na Seção \ref{sec:dados.observacionais.ogle.iii.resultados}, são discutidos os dados fotométricos das estrelas Variáveis Cefeidas disponíveis no catálogo OGLE-IV. Realiza-se a análise desses dados, incluindo a discussão do algoritmo computacional "Periodograma Lomb-Scargle". São obtidos os períodos de pulsação das estrelas Variáveis Cefeidas na Grande Nuvem de Magalhães e, após análise e remoção dos dados que não foram possíveis de serem tratados pelo Periodograma Lomb-Scargle, calcula-se a Lei de Leavitt para a magnitude aparente das Variáveis Cefeidas da Grande Nuvem de Magalhães.

Na Seção \ref{sec:determinando.distancia.grande.nuvem.magalhaes.via.lactea}, desenvolvemos uma metodologia para calibrar a distância até a Grande Nuvem de Magalhães. Utilizamos a Lei de Leavitt e o módulo de distância para estabelecer uma relação entre a magnitude média e o período das Variáveis Cefeidas, divididas em duas regiões distintas no diagrama período-luminosidade. Introduzimos a grandeza \(\delta \mu\), representando a variação na luminosidade intrínseca das estrelas, e derivamos uma fórmula para calcular a distância com base no período e na magnitude aparente média dessas estrelas. Validamos o método usando uma distância de calibração obtida pelo método das binárias eclipsantes (com precisão de 1,27\%) \cite{pietrzynski2019distance}, garantindo a confiabilidade da determinação da distância para a Região I (\( (50,57 \pm 0,91) \text{k}\,\text{pc} \)) e para a Região II (\( (51,80 \pm 0,85) \text{k}\,\text{pc} \)).

\section{Modelo de Pulsação Estelar}
\label{sec:modelo.pulsacao.estelar}
\begin{figure}[!h]
	\centering
	\includegraphics[scale=0.5]{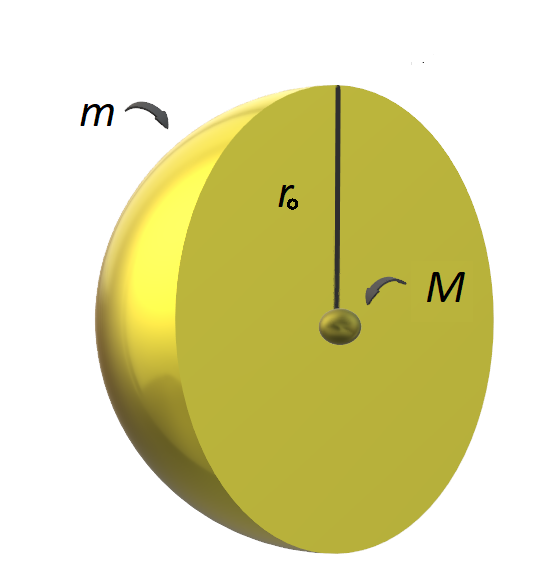}
	\caption{Protótipo de pulsação estelar. Criado por um dos autores deste trabalho (KMC).}
	\label{fig:prototipo.estelar}
\end{figure}

Com o intuito de dar uma breve explicação teórica sobre as estrelas variáveis, é oportuno discorrer sobre o seu mecanismo de pulsação. Definiremos, então, um modelo simplificado de estrela variável, cuja variação ocorre devido ao balanço gravitacional e termodinâmico da estrela. Imaginemos uma estrela esférica cuja massa $M$ está totalmente concentrada em seu centro, essa estrela possui uma casca atmosférica de massa $m$ à uma distância $r_0$ do centro e possui uma pressão interna $P_0$ (veja na Figura \ref{fig:prototipo.estelar}). Suponha que a estrela esteja imersa no vácuo (Logo, a pressão exterior à casca atmosférica é nula) \cite{Tanner2002}.

 Aplicando a segunda lei de Newton sobre a casca atmosférica:
\begin{equation}
    m \frac{d^2 R}{dt^2} = - \frac{G M m}{R^2} + 4 \pi R^2 P.
    \label{eq: dinori}
\end{equation}
\par Assume-se que $R = r_0 + \delta r$ é o raio perturbado da estrela  e $P = P_0 + \delta p$ a pressão perturbada da estrela, 
\begin{equation}
    m \frac{d^2 (r_0 + \delta r)}{dt^2} = - \frac{G M m}{(r_0 + \delta r)^2} + 4 \pi (r_0 + \delta r)^2 (P_0 + \delta p).
    \label{eq:newton}
\end{equation}
\par Aplicando a aproximação de primeira ordem:
\begin{equation}
    \frac{1}{(r_0 + \delta r)^2} \approx \frac{1}{{r_0}^2} \left( 1 - 2 \frac{\delta r}{r_0}  \right).
\end{equation}
\par Se negligenciarmos os termos de ordem superior de $\delta p$ a Equação (\ref{eq:newton}) toma a forma:
\begin{equation}
    m \frac{d^2 (\delta r)}{dt^2} = - \frac{G M m}{r_0^2} + \frac{2 G M m \delta r}{r_0^3} + 4 \pi {r_0}^2 P_0 + 8 \pi r_0 P_0 \delta r + 4 \pi {r_0}^2 \delta p, 
    \label{eq:dinamica}
\end{equation}
\par a solução para o equilíbrio é:
\begin{equation}
    - \frac{G M m}{r_0^2} + 4 \pi {r_0}^2 P_0 = 0 .
    \label{eq: equilibrio}
\end{equation}

\par Aplicando a solução de equilíbrio na Equação (\ref{eq:dinamica}), temos:

\begin{equation}
    m \frac{d^2 (\delta r)}{dt^2} = \frac{2 G M m \delta r}{r_0^3} + 8 \pi r_0 P_0 \delta r + 4 \pi {r_0}^2 \delta P,
    \label{eq:dinam}
\end{equation}
se assumirmos a pulsação estelar como uma contração e expansão de um gás adiabático de coeficiente adiabático $\gamma$, tem-se,

\begin{equation}
    P_0 {V_0}^\gamma = P V^\gamma,
    \label{eq:pv}
\end{equation}
o que implica em,
\begin{equation}
    P {V}^\gamma = \text{constante}.
\end{equation}

Visto que o volume da estrela é dado ela expressão:

\begin{equation}
    V = \frac{4}{3} \pi R^3,
    \label{eq:vol}
\end{equation}
então tem-se que,
\begin{equation}
    P R^{3 \gamma} = \text{constante}.
\end{equation}
\par Portanto: 
\begin{equation}
    \frac{\delta P}{P_0} = - 3 \gamma \frac{\delta r}{r_0}.
    \label{eq:term}
\end{equation}

\par Substituindo (\ref{eq: equilibrio}) e (\ref{eq:term}) em (\ref{eq:dinam}):
\begin{equation}
    \frac{d^2 \delta r}{dt^2} = - (3\gamma - 4) \frac{GM}{{r_0}^3} \delta r.
\end{equation}
\par Se $\gamma > \frac{4}{3}$, a solução para a perturbação é do tipo senoidal com período de pulsação, 
\begin{equation}
    T = \displaystyle\frac{2 \pi}{ \sqrt{(3 \gamma - 4) \frac{GM}{{r_0}^3}}}.
\end{equation}

\par Para um gás ideal $\gamma = \frac{5}{3}$ \cite{salinas2008}, então - assumindo uma estrela formada por um gás ideal - o período da solução perturbativa de primeira ordem será:
\begin{equation}
    T_\text{pertubado} = {2 \pi}{\sqrt{\frac{{r_0}^3}{GM}}}.
\end{equation}

\par Por meio da abordagem perturbativa, compreendemos de forma concisa as oscilações das estrelas variáveis, identificando estrelas com períodos de pulsação bem definidos. No entanto, para além dessa solução perturbativa, é possível investigar a natureza oscilatória do modelo por meio da solução numérica da Equação (\ref{eq: dinori}). Se usarmos a Equação (\ref{eq:pv}) e a Equação (\ref{eq:vol}), temos:

\begin{equation}
    P = \frac{P_0}{R^5}.
\end{equation}
\par Assumindo $R = r_0 r$, se obtém a equação:
\begin{equation}
    \frac{d^2 r}{dt^2} = -\frac{GM}{{r_0}^3 r^2} +  \frac{4\pi r_0 P_0}{m r^3} .
    \label{eq:dianmo}
\end{equation}

A Equação (\ref{eq:dianmo}) pode ser reescrita na forma:
\begin{equation}
    \frac{d^2 r}{dt^2} = \frac{A}{r^3} -  \frac{B}{r^2}.  
    \label{eq:casca}
\end{equation}

\par
\begin{figure}[H]
	\centering
	\includegraphics[scale=0.5]{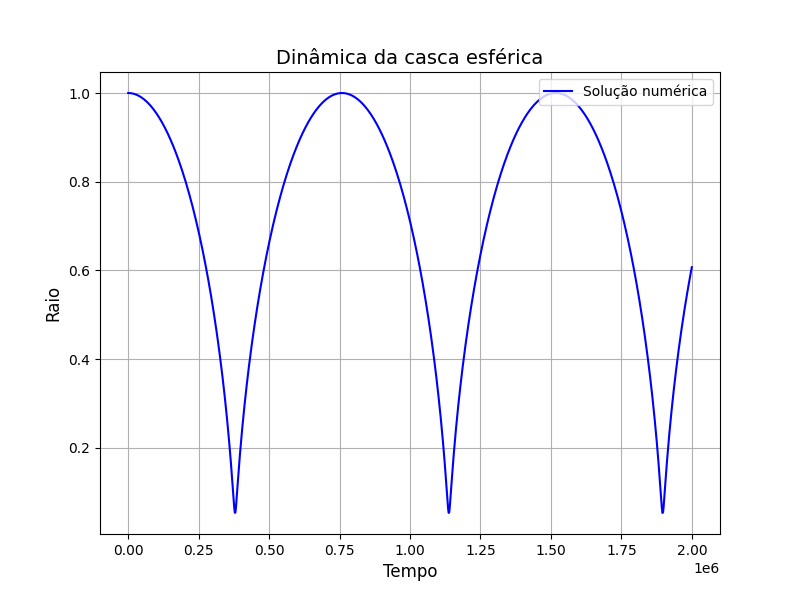}
	\caption{Solução numérica para a Equação (\ref{eq:casca}). Criado por um dos autores deste trabalho (KMC).}
	\label{fig:numerica}
\end{figure}

 Tomando as condições iniciais $r(t=0) = 1.0$ e $\left. \frac{d r}{dt} \right|_{t=0} = 0.0$ e os valores de $A = 10^{-12}$ e $B = 10^{-11}$ se obtém a curva ilustrada na Figura \ref{fig:numerica} 

\section{Dados observacionais OGLE-IV e resultados}
\label{sec:dados.observacionais.ogle.iii.resultados}
\subsection{O catálogo OGLE-IV}

Existem diversos projetos que fornecem dados sobre estrelas Variáveis Cefeidas. Neste estudo, optamos por utilizar exclusivamente o catálogo \textit{"The Optical Gravitational Lensing Experiment Collection of Variable Stars"}. Este projeto está em operação desde 1992, coletando dados observacionais tanto das Nuvens de Magalhães quanto da Via Láctea, monitorando mais de 200 milhões de estrelas. As observações foram realizadas com os filtros Cousins I e Johnson V, mas no presente trabalho foi adotado unicamente o filtro Cousins I, que corresponde ao infravermelho próximo com um comprimento de onda médio de $789n\:m$ e uma largura de banda de $154 n\: m$. As observações são conduzidas com um tempo de exposição de aproximadamente 180 segundos e em média são realizadas cerca de 400 medidas. Os dados são disponibilizados no formato DAT, incluindo informações sobre a magnitude aparente e o tempo (em Dias Julianos) \cite{udalski2015ogle}. Os dados da 4ª fase do projeto (OGLE–IV) são de acesso público. Esta é a fonte de dados que adotamos neste trabalho.

\par
\begin{figure}[H]
	\centering
	\includegraphics[scale=0.5]{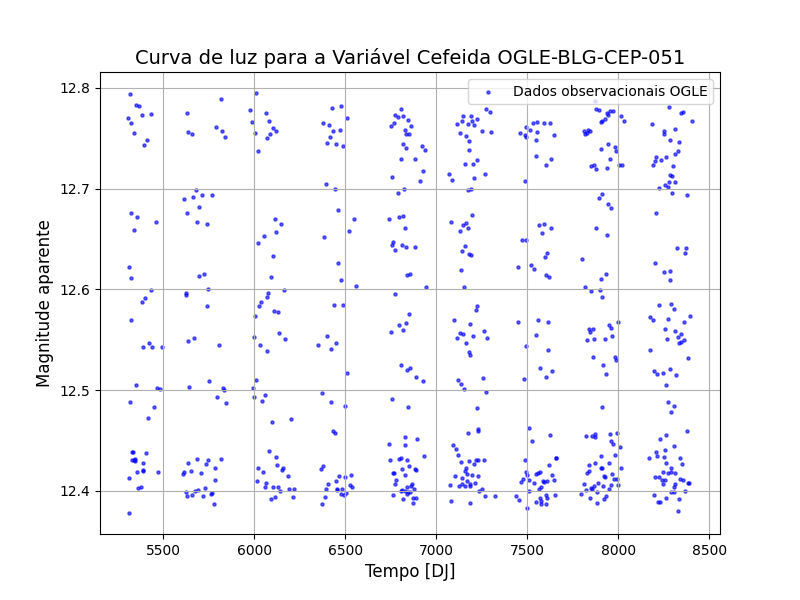}
	\caption{Curva de luz para a Variável Cefeida OGLE-BLG-CEP-051 localizada no Bojo da Via Láctea obtida do catálogo OGLE-IV. Criada por um dos autores deste trabalho (KMC).}
	\label{fig:curva.luz.def}
\end{figure}

 Devido ao grande volume de observações para cada estrela, o processo de determinar visualmente o período das estrelas torna-se altamente complexo, isso ocorre porque a análise manual de tal quantidade de dados é suscetível a erros e imprecisões , como ilustrado na Figura \ref{fig:curva.luz.def}. Isso levanta a necessidade de um tratamento computacional dos dados.\\

\subsection{Analise dos dados}
\label{sec:analise}

\subsubsection{O Periodograma Lomb-Scargle}
\label{sec:Periodograma.lomb-scargle}

Antes de adentrarmos na explicação do algoritmo empregado para calcular a distância, é pertinente iniciar com uma breve introdução à ferramenta denominada "Periodograma". Esta ferramenta desempenha um papel fundamental na análise espectral de sinais e séries temporais, sendo frequentemente empregada em diversas áreas científicas, tais como processamento de sinais, meteorologia, engenharia elétrica, econometria, etc \cite{torrence1998practical}.

O Periodograma de Schuster, que foi originalmente proposto por Schuster em 1898, representa uma das abordagens clássicas para calcular o Periodograma. A essência desse Periodograma reside na identificação de padrões ou periodicidades presentes nos dados temporais \cite{schuster1898investigation}.

A principal equação que constitui o núcleo do Periodograma é uma expressão matemática que quantifica a distribuição da energia espectral em diversas frequências presentes nos dados. Essa fórmula é precisamente definida da seguinte forma:

\begin{equation}
    P(f) = \frac{1}{N} \sum_{n=1}^{N} g_n e^{-2\pi i k n/N} ,
\end{equation}  onde:

- \(P(f)\): Representa a potência.

- \(N\): É o número de medidas ou pontos de dados na série temporal.

- \(g_n\): São os valores dos dados amostrados em diferentes momentos no tempo.

- \(k\): É um índice que varia de 0 a \(N-1\) e está associado às frequências. Quanto maior o valor de \(k\), maior a frequência em consideração.

O resultado da computação do Periodograma é uma função que descreve a distribuição da energia espectral em diversas frequências. Ao examinarmos o Periodograma, podemos identificar picos, isto é, frequências nas quais a energia se encontra consideravelmente amplificada. Estes picos sinalizam as frequências predominantes nos dados.

É crucial notar que o Periodograma de Schuster parte da premissa de que os dados estão distribuídos de forma equidistante no tempo, o que pode não ser o caso em muitas situações do mundo real. Para abordar conjuntos de dados que apresentam intervalos irregulares entre as observações, foi desenvolvida uma técnica correlata denominada Periodograma de Lomb-Scargle. Esta técnica foi concebida para acomodar essa irregularidade nos dados e oferecer uma análise espectral precisa em tais cenários \cite{lomb1976leastsquares, scargle1982studies, barning1962numerical}.

\begin{figure}[H]
	\centering
	\includegraphics[scale=0.5]{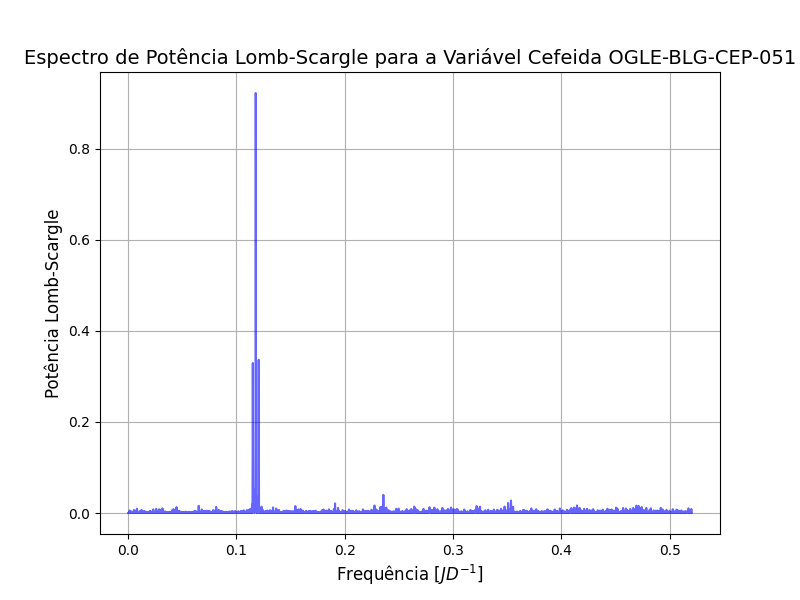}
	\caption{Espectro de potência Lomb-Scargle para a Variável Cefeida OGLE-BLG-CEP-051 do catálogo OGLE-IV com pico de potência Lomb-Scargle associado à frequência de $0.118\; {JD}^{-1}$ (Equivalente a um período de $8.49$ Dias Julianos), obtida a partir de um algoritmo criado por um dos autores deste trabalho (KMC).}
	\label{fig:espectro}
\end{figure}

Nesse trabalho foi adotado o Periodograma Lomb-Scargle, e a versão do algoritmo Lomb-Scargle empregada neste trabalho é a função \texttt{LombScargle} do módulo \texttt{timeseries} da biblioteca Astropy na linguagem de programação Python 3.9. Essa função comporta dois vetores de mesmo tamanho: um contendo uma sequência de dados temporais e o outro contendo dados relacionados ao tempo. Ela fornece dois vetores, um com as frequências e outro com o espectro de potência para cada frequência. O espectro de potência, representado na Figura \ref{fig:espectro}, indica a intensidade da variabilidade nas diferentes frequências presentes nos dados \cite{astropy_lombscargle}.

\par
\begin{figure}[H]
	\centering
	\includegraphics[scale=0.5]{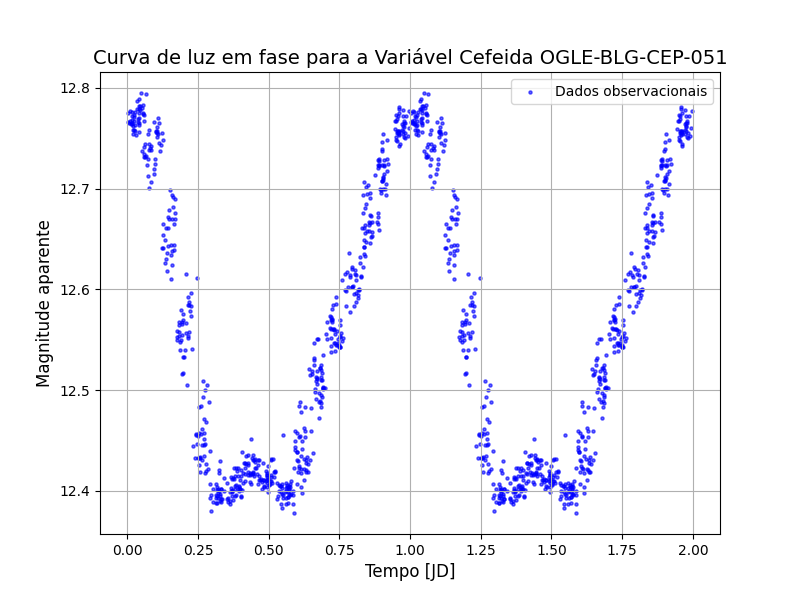}
	\caption{Curva de luz em fase com o período de 8,49 Dias Julianos para a Variável Cefeida OGLE-BLG-CEP-051 do catálogo OGLE-IV, obtida a partir de um algoritmo criado por um dos autores deste trabalho (KMC).}
	\label{fig:curva.luz.fase}
\end{figure}

 A interpretação desse resultado é que o pico mais alto no espectro Lomb-Scargle indica uma frequência predominante nos dados, e o período correspondente a essa frequência é uma estimativa do período dominante \cite{astropy_lombscargle}. Ao aplicar o período Lomb-Scargle aos dados da Variável Cefeida OGLE-BLG-CEP-051 (representada na Figura \ref{fig:curva.luz.def}) do catálogo OGLE-IV, obtemos a curva exibida na Figura \ref{fig:curva.luz.fase}.

\subsubsection{Algoritmo para a determinação das distâncias}
\label{sec:algoritmo.distancia}

Com o intuito de analisar os dados das Variáveis Cefeidas do catálogo OGLE-IV, desenvolvemos um algoritmo que passa por quatro etapas: carregamento e limpeza de dados, cálculo de períodos, obtenção do espaço de fase e geração das curvas de luz. O objetivo primordial é estabelecer uma Lei de Leavitt precisa e confiável para análises posteriores.

É importante observar que o algoritmo foi projetado para operar em um diretório principal que contenha um subdiretório adicional denominado "DATABASE". Este subdiretório deve conter arquivos de texto individuais para cada estrela, organizados em três colunas distintas: a primeira coluna registra a data em Dias Julianos das observações, enquanto a segunda coluna registra as magnitudes aparentes medidas nessa data. A essa estrutura é fundamental para assegurar o funcionamento adequado do algoritmo, possibilitando uma análise precisa das Variáveis Cefeidas e contribuindo para a acurácia dos resultados neste estudo.\\

\par O algoritmo é dividido em quatro etapas principais, cada uma com seus objetivos específicos.

\begin{enumerate}
    \item \textbf{Primeira Etapa:} Nesta fase inicial, o algoritmo realiza o carregamento dos arquivos necessários e realiza a limpeza da memória, assegurando que não haja resquícios de processamentos anteriores que possam afetar a análise atual.
    
    \item \textbf{Segunda Etapa:} A segunda etapa é dedicada ao cálculo dos períodos das estrelas. Isso é realizado por meio do método do Periodograma Lomb-Scargle, conforme detalhado na Seção \ref{sec:Periodograma.lomb-scargle}. Essa etapa é crucial para determinar o período de oscilação das Variáveis Cefeidas.
    
    \item \textbf{Terceira Etapa:} Com os períodos calculados na etapa anterior, o algoritmo avança para o cálculo do espaço de fase (configuração na qual o tempo é particionados e sobreposto em intervalos periódicos, como será exemplificado a seguir) das estrelas. Essa etapa é essencial para criar uma representação visual das oscilações das Variáveis Cefeidas, facilitando a análise posterior.
    
    \item \textbf{Quarta Etapa:} Na última etapa, o algoritmo gera a curva de luz com base no espaço de fase calculado na etapa anterior. Além disso, quantifica a dispersão das curvas de luz, aplica um teste de dispersão e remove as estrelas que não atendem aos critérios estabelecidos. Finalmente, a Lei de Leavitt é calculada com base nas estrelas que passaram no teste de dispersão.
\end{enumerate}

A parte fundamental do algoritmo está nas etapas segunda, terceira e quarta. Portanto, estas etapas serão discutidas com mais detalhes. No entanto, no GitHub, há uma explicação completa sobre como o algoritmo funciona \cite{costa2023distanciasgalacticas}.

Na segunda etapa, ocorre o cálculo do período de oscilação das estrelas. Após a conclusão desses cálculos, os dados de período-luminosidade são plotados, e um ajuste é realizado usando a função \texttt{curve\_fit} do módulo \texttt{optimize} da biblioteca \texttt{scipy}. Esse ajuste é aplicado a dados seguindo uma função da forma

\begin{equation} \label{eq:mod.leavitt}
    m = a \log(P) + b.
\end{equation}

\par A função \texttt{curve\_fit} possibilita o cálculo do erro associado ao ajuste, por meio de uma matriz de covariância dos parâmetros do ajuste com o modelo adotado. Esse erro está relacionado às incertezas na Lei de Leavitt, conforme as Equações \ref{eq:leavitt-I} e \ref{eq:leavitt-II}.

\par Após a conclusão da etapa, obtém-se a Lei de Leavitt. No entanto, é importante observar que nem todos os períodos calculados são precisos, e nem todos os dados estão em total conformidade com a Lei de Leavitt. Os dados adquiridos na natureza podem conter erros estatísticos e sistemáticos em suas medições, além de a natureza não se comportar exatamente de acordo com as expectativas teóricas.

Portanto, torna-se necessário realizar um estudo individual das estrelas, bem como uma análise global da Lei de Leavitt obtida. Esses estudos ocorrem na quarta etapa, utilizando os resultados da terceira etapa.\\

\par Na terceira etapa são calculadas as curvas de luz, para isso utilizam-se os dados $(t,m)$ do tempo e da magnitude aparente, os dados são particionados periodicamente em dados do tempo $t$ (utilizando o período calculado para a estrela) e se calculam as partições dos dados de tempo sobrepostas, normalizando o período de pulsação para $P=1$, esse procedimento é similar ao procedimento de ajustar cada conjunto de dados $(P_i , m_i)$ em uma função periódica $m_i = f(t,\omega , \phi_i)$ - para cada conjunto de dados $i$ - e ajustar a fase $\phi_i$ de cada conjunto a fim de sobrepor as funções, esse procedimento foi denominado "obter a curva de luz em fase".\\

\par
\begin{figure}[H]
    \centering
    \begin{minipage}{0.54\textwidth}
        \centering
        \includegraphics[scale=0.5]{figura/curva-de-luz-fase.png}
    \end{minipage}%
    \begin{minipage}{0.5\textwidth}
        \centering
        \includegraphics[scale=0.5]{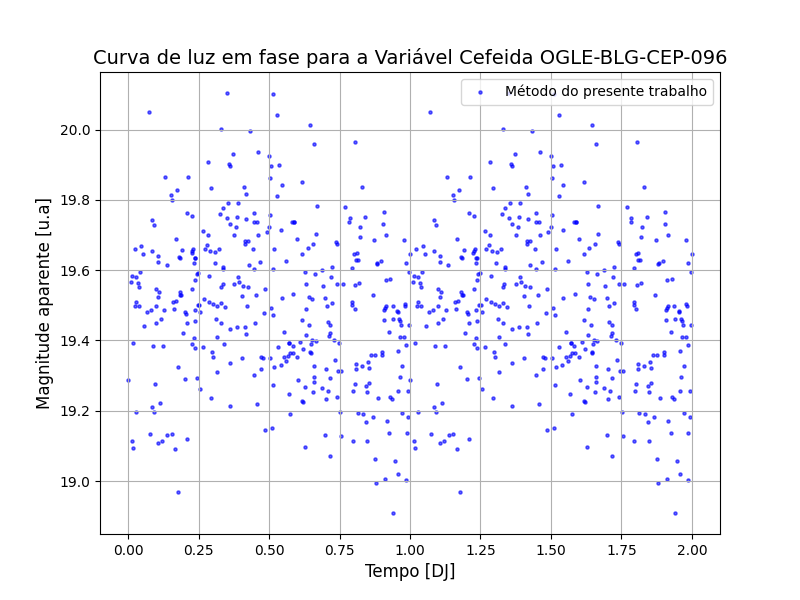}
    \end{minipage}
    \caption{Exemplo da região completa utilizada para o cálculo de dispersão para a estrela OGLE-BLG-CEP-051 e para a estrela OGLE-BLG-CEP-096 obtida a partir da metodologia discutida na Seção \ref{sec:algoritmo.distancia}. Criada por um dos autores deste trabalho (KMC).}
    \label{fig:complete}
\end{figure}

\begin{figure}[H]
    \centering
    \begin{minipage}{0.5\textwidth}
        \centering
        \includegraphics[scale=0.5]{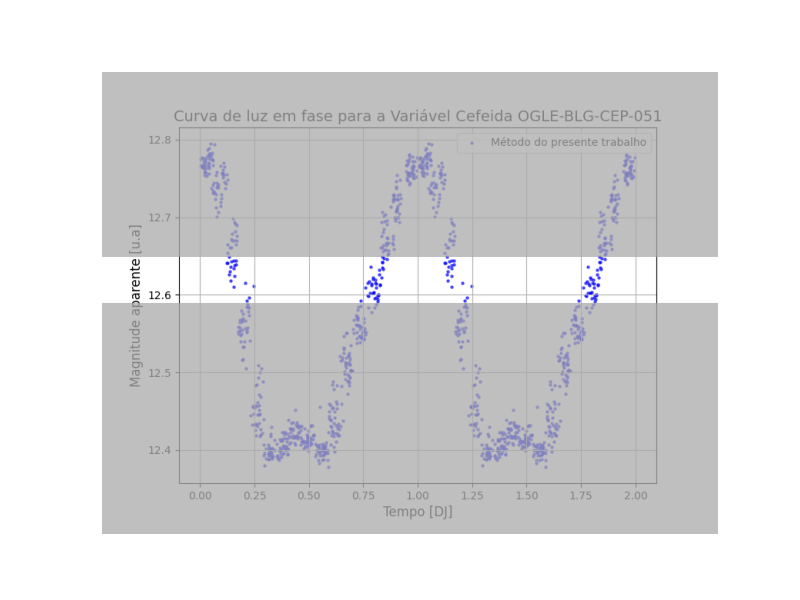}
    \end{minipage}%
    \begin{minipage}{0.5\textwidth}
        \centering
        \includegraphics[scale=0.5]{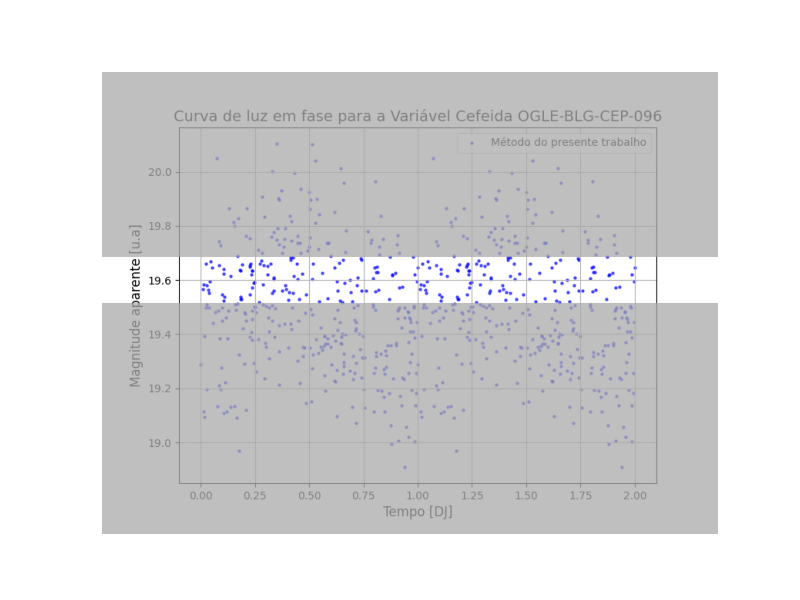}
    \end{minipage}
    \caption{Exemplo da região horizontal utilizada para o cálculo de dispersão para a estrela OGLE-BLG-CEP-051 e para a estrela OGLE-BLG-CEP-096 obtida a partir da metodologia discutida na Seção \ref{sec:algoritmo.distancia}.Criada por um dos autores deste trabalho (KMC).}
    \label{fig:horizontal}
\end{figure}

\begin{figure}[H]
    \centering
    \begin{minipage}{0.5\textwidth}
        \centering
        \includegraphics[scale=0.5]{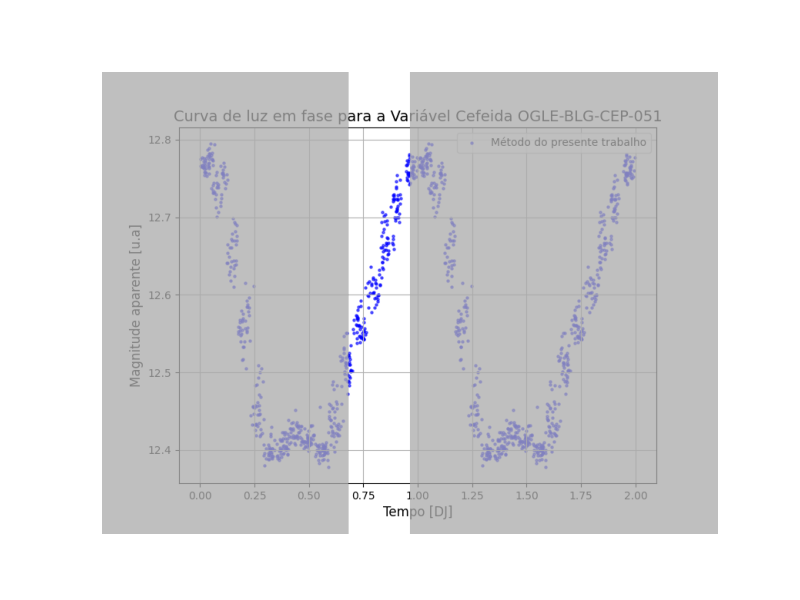}
    \end{minipage}%
    \begin{minipage}{0.5\textwidth}
        \centering
        \includegraphics[scale=0.5]{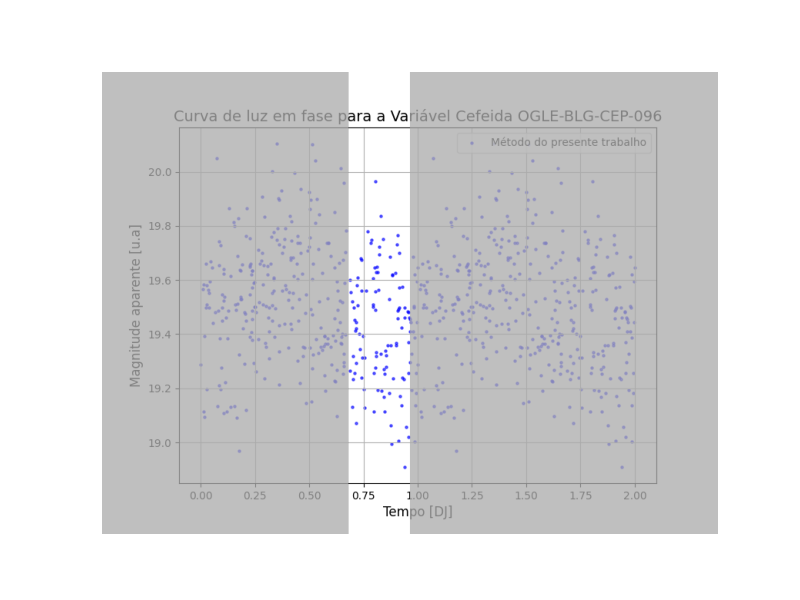}
    \end{minipage}
    \caption{Exemplo da região vertical utilizada para o cálculo de dispersão para a estrela OGLE-BLG-CEP-051 e para a estrela OGLE-BLG-CEP-096 obtida a partir da metodologia discutida na Seção \ref{sec:algoritmo.distancia}. Criada por um dos autores deste trabalho (KMC).}
    \label{fig:vertical}
\end{figure}

 Na quarta etapa, são geradas as curvas de luz em fase, as quais foram calculadas na terceira etapa. Neste estágio, é estabelecido um coeficiente de dispersão que quantifica a proporção de pixels ciano e pixels brancos presentes na imagem analisada, que representa a curva de luz da estrela em fase. Essa quantificação é determinada calculando a razão entre os pixels ciano e os pixels brancos na imagem gerada e é realizada em três áreas distintas da imagem analisada, a saber: a imagem completa (Figura \ref{fig:complete}), um corte horizontal no centro da imagem (Figura \ref{fig:horizontal}) e um corte vertical no centro da imagem (Figura \ref{fig:vertical}).


\par As imagens que não atenderem aos critérios dos testes (que serão detalhados posteriormente) têm seus dados correspondentes, localizados no diretório "DATABASE", transferidos para um diretório chamado "DISPERSAO". É fundamental que o diretório "DISPERSAO" seja criado manualmente pelo usuário, garantindo que o programa não tenha acesso a esses dados que foram rejeitados.

\begin{figure}[H]
	\centering
	\includegraphics[scale=0.5]{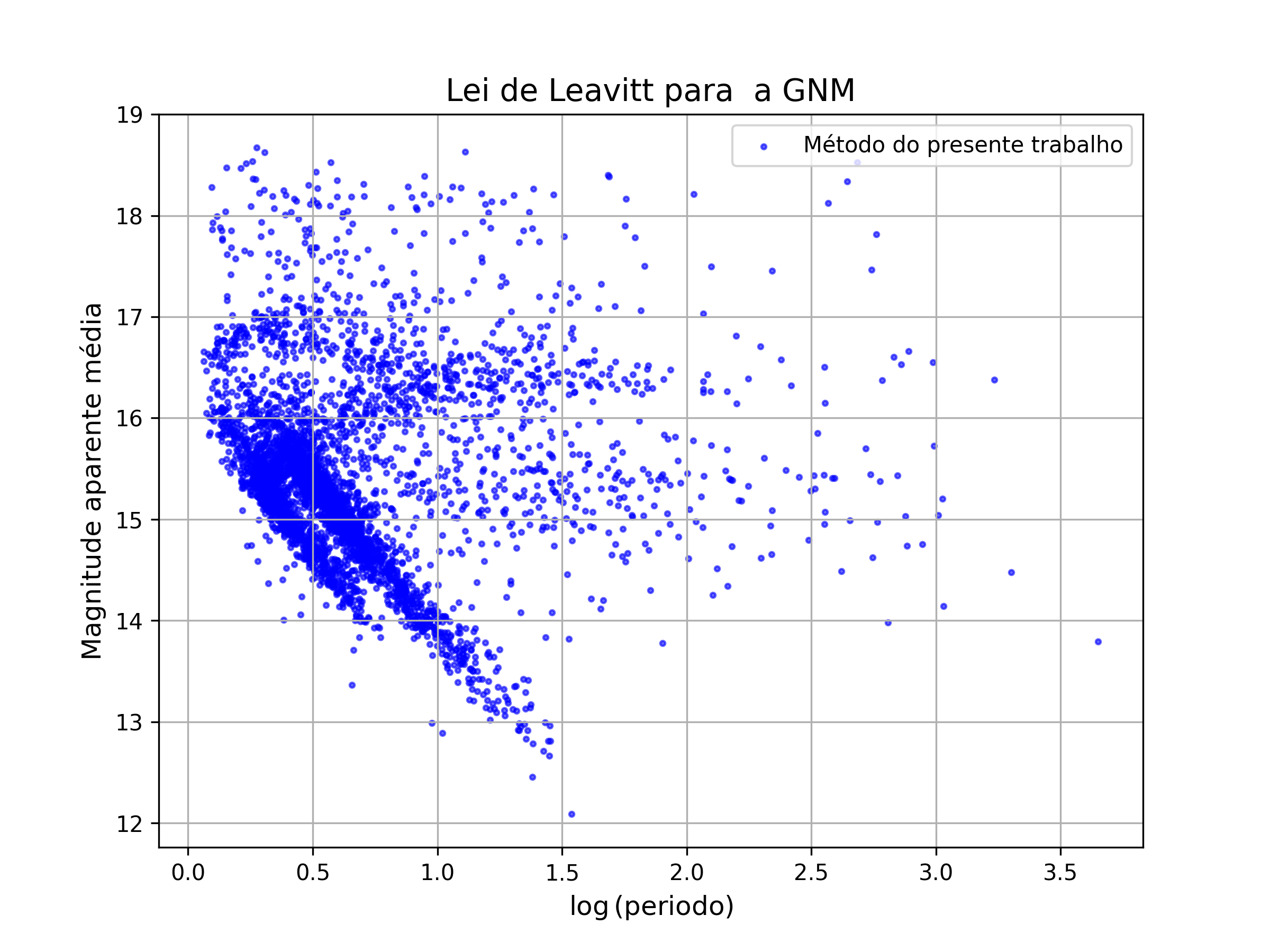}
	\caption{Lei de Leavitt para a Grande Nuvem de Magalhães antes do teste de dispersão, usando as Variáveis Cefeidas do catálogo OGLE-IV, obtida a partir do algoritmo criado por um dos autores deste trabalho (KMC).}
	\label{fig:leavitt.antes.lmc}
\end{figure}

\begin{figure}[H]
	\centering
	\includegraphics[scale=0.5]{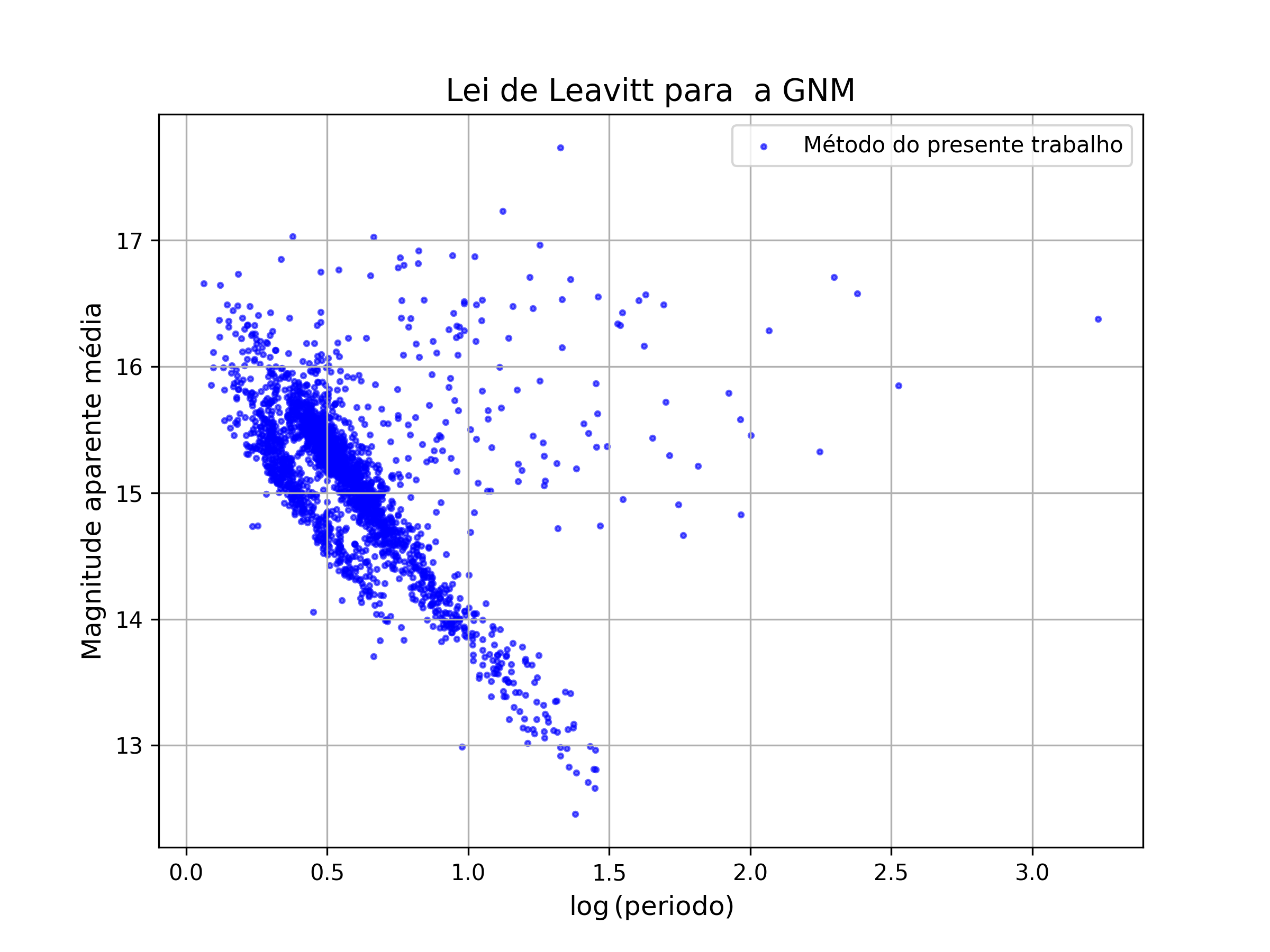}
	\caption{Lei de Leavitt para a Grande Nuvem de Magalhães depois do teste de dispersão, usando as Variáveis Cefeidas do catálogo OGLE-IV, obtida a partir do algoritmo criado por um dos autores deste trabalho (KMC).}
	\label{fig:leavitt.depois.lmc}
\end{figure}

Para avaliar a qualidade das imagens, foram selecionadas algumas estrelas cujos períodos estavam bem definidos. Posteriormente, os coeficientes de dispersão foram calculados para as três regiões mencionadas anteriormente. Com base nos coeficientes calculados, foi construído um intervalo de confiança no qual se espera encontrar as estrelas com os períodos definidos mais próximos do ideal. As estrelas que se encontraram fora desse intervalo de confiança foram rejeitadas no teste. As Figuras \ref{fig:leavitt.antes.lmc} e \ref{fig:leavitt.depois.lmc} exemplificam o diagrama período-luminosidade para a Grande Nuvem de Magalhães antes e após a realização do teste.

\par
\begin{figure}[H]
	\centering
	\includegraphics[scale=0.5]{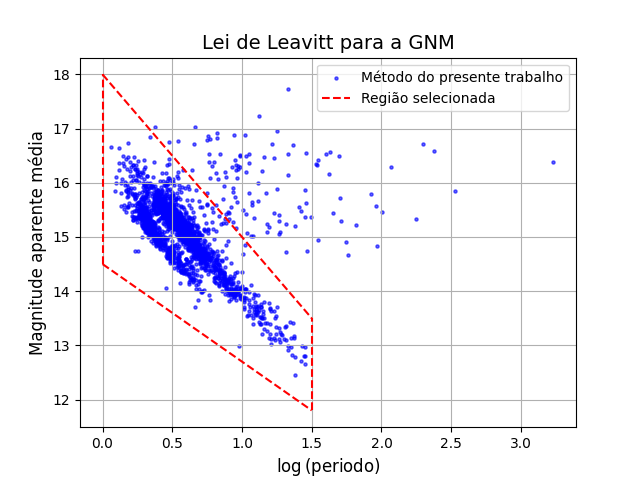}
	\caption{Seleção das estrelas Cefeidas da Grande Nuvem de Magalhães resultantes do teste de dispersão, usando as Variáveis Cefeidas do catálogo OGLE-IV, obtida a partir do algoritmo criado por um dos autores deste trabalho (KMC).}
	\label{fig:regiao-antes-lmc}
\end{figure}

\begin{figure}[H]
	\centering
	\includegraphics[scale=0.5]{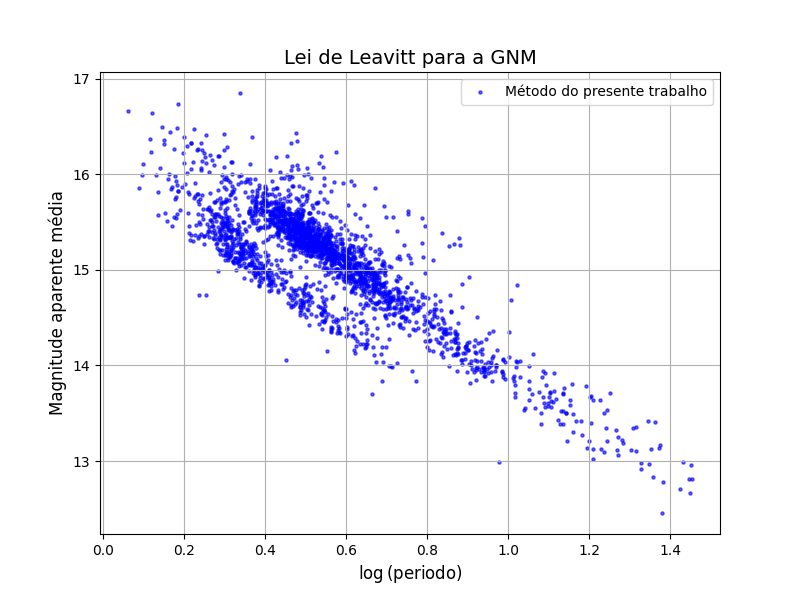}
	\caption{Lei de Leavitt para a Grande Nuvem de Magalhães, usando as Variáveis Cefeidas do catálogo OGLE-IV, a partir do algortimo criado por um dos autores deste trabalho (KMC).}
	\label{fig:curva.luz.lmc}
\end{figure}

 Após essa análise, observa-se que não foi possível filtrar todas as estrelas que não se comportam de acordo com a lei de Leavitt. Para obter os resultados representados na Figura \ref{fig:curva.luz.lmc}, foi necessário realizar a remoção manual das estrelas que passaram no teste de qualidade, mas ainda exibiam uma dispersão significativa no diagrama período-luminosidade. A área selecionada para essa remoção está indicada na Figura \ref{fig:regiao-antes-lmc}.

\par Depois da seleção, chegamos a um diagrama período-luminosidade aparente, que representa a Lei de Leavitt das estrelas da Grande Nuvem de Magalhães (Figura \ref{fig:curva.luz.lmc}).

\begin{figure}[H]
	\centering\includegraphics[scale=0.5]{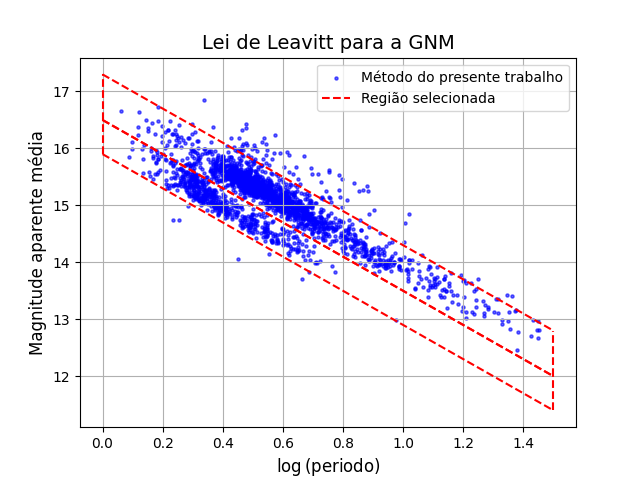}
	\caption{Lei de Leavitt para diferentes regiões do diagrama período-luminosidade das estrelas Cefeida da Grande Nuvem de Magalhães do catálogo OGLE-IV, a partir do algoritmo criado por um dos autores deste trabalho (KMC).}
	\label{fig:regioes}
\end{figure}

Na Figura \ref{fig:curva.luz.lmc}, é evidente que a galáxia apresenta dois comportamentos distintos na relação período-luminosidade de suas estrelas (consulte a Figura \ref{fig:regioes}). Isso ocorre devido aos diferentes modos de oscilação das Variáveis Cefeidas, conhecidos como modo fundamental e primeiro sobreton. No "modo fundamental", a estrela expande e contrai de maneira uniforme durante seu ciclo de pulsação, e sua luminosidade está diretamente relacionada ao período de pulsação, sendo que estrelas com períodos mais longos são mais luminosas. Em contrapartida, no "primeiro sobreton", a estrela pulsa de maneira mais complexa, com diferentes camadas expandindo e contraindo de forma desigual, resultando em uma relação período-luminosidade diferente. Cefeidas em modo primeiro sobreton são mais luminosas para um dado período de pulsação em comparação com aquelas no modo fundamental com o mesmo período \cite{most-cepheids}.

\subsection{Comparação com os resultados da colaboração OGLE-IV}

\begin{figure}[H]
\centering
\begin{subfigure}{0.45\textwidth}
  \centering
  \includegraphics[width=\linewidth]{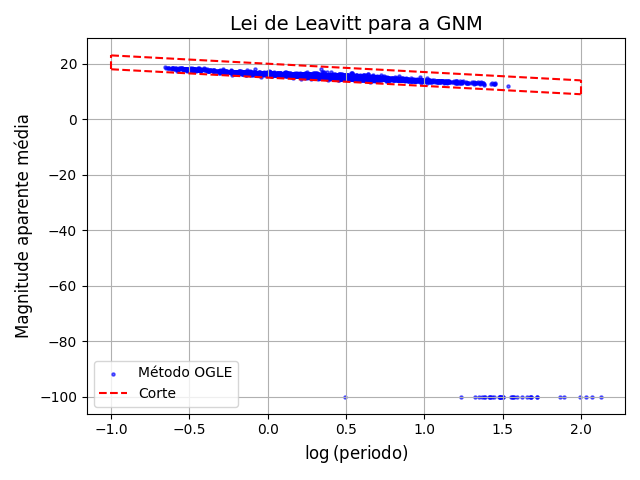}
  \label{fig:sub1}
\end{subfigure}%
\hspace{0.05\textwidth}
\begin{subfigure}{0.45\textwidth}
  \centering
  \includegraphics[width=\linewidth]{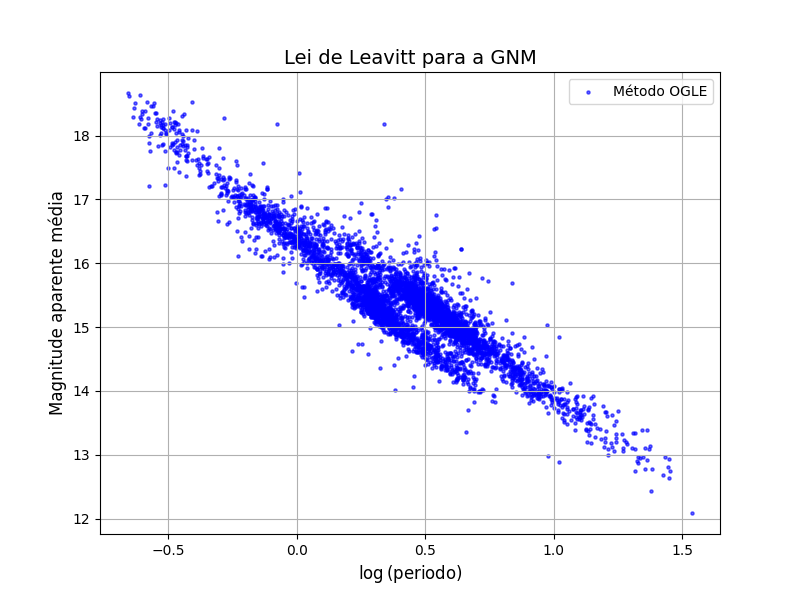}
  \label{fig:sub2}
\end{subfigure}
\caption{Seleção das Variáveis Cefeidas da Grande Nuvem de Magalhães (À esquerda), e Lei de Leavitt para a Grande Nuvem de Magalhães para os resultados obtidos pela colaboração OGLE-IV (À direita). obtida a partir do resultado da colaboração OGLE-IV.}
\label{fig:combined}
\end{figure}

O catálogo OGLE-IV também disponibiliza o período de oscilação das estrelas Variáveis Cefeidas em sua base de dados. Portanto, é de interesse realizar uma comparação entre os resultados obtidos pelo projeto OGLE-IV e os resultados deste trabalho. Tanto para os resultados obtidos pela colaboração OGLE-IV quanto para os resultados deste trabalho, foi necessária a seleção manual das estrelas que mantiveram uma dispersão significativa em relação à região mais densamente povoada. A área escolhida para essa seleção está representada na Figura \ref{fig:combined}.

\begin{figure}[H]
	\centering
	\includegraphics[scale=0.5]{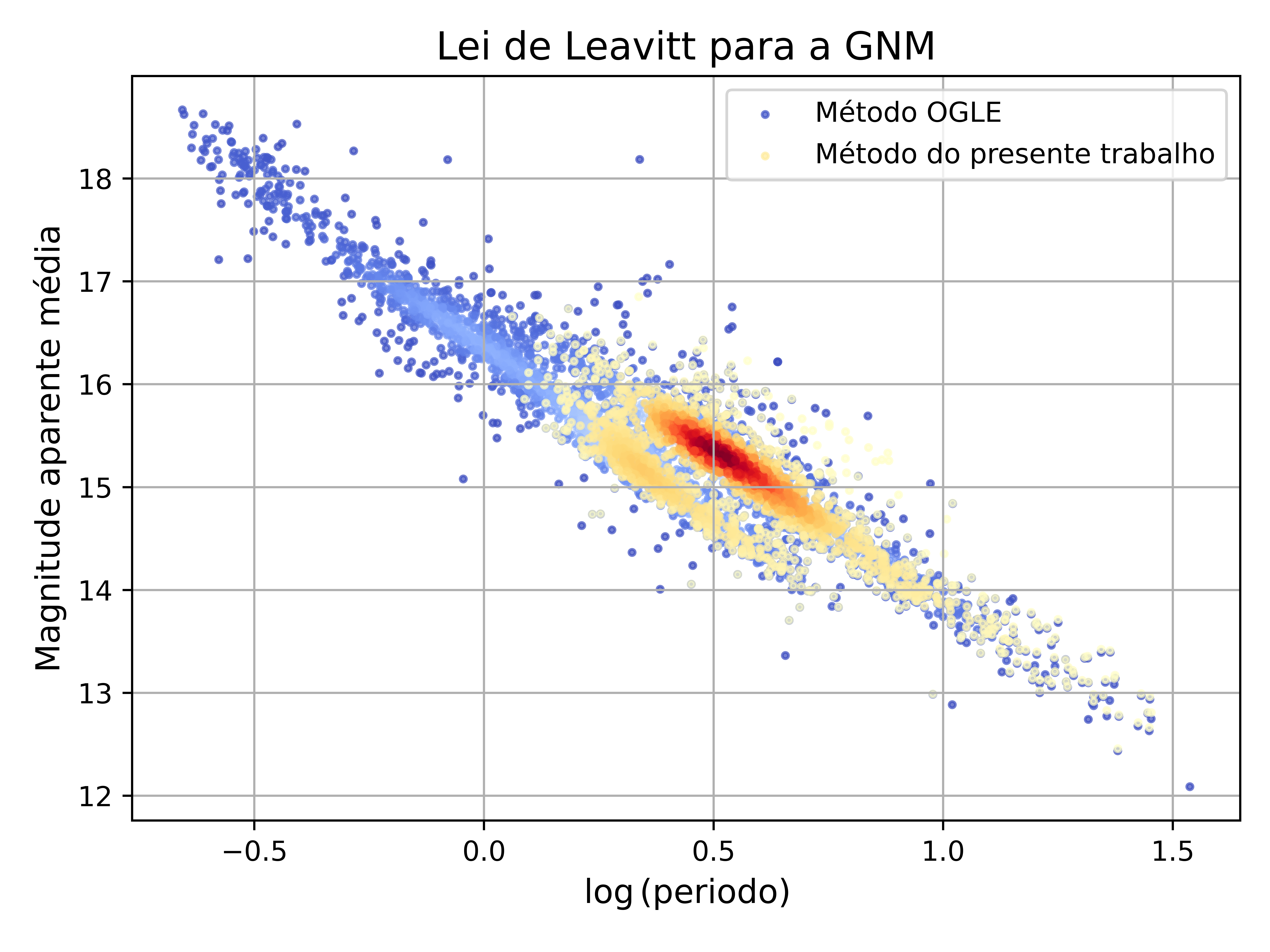}
	\caption{Sobreposição da Lei de Leavitt para a Grande Nuvem de Magalhães com os resultados obtidos pela colaboração OGLE-IV e por este trabalho.  Criado por um dos autores deste trabalho (KMC).}
	\label{fig:sobreposicao}
\end{figure}

O diagrama com a sobreposição do diagrama período-luminosidade obtido por ambos os trabalhos está representado na Figura \ref{fig:sobreposicao}.

\begin{figure}[H]
	\centering
	\includegraphics[scale=0.5]{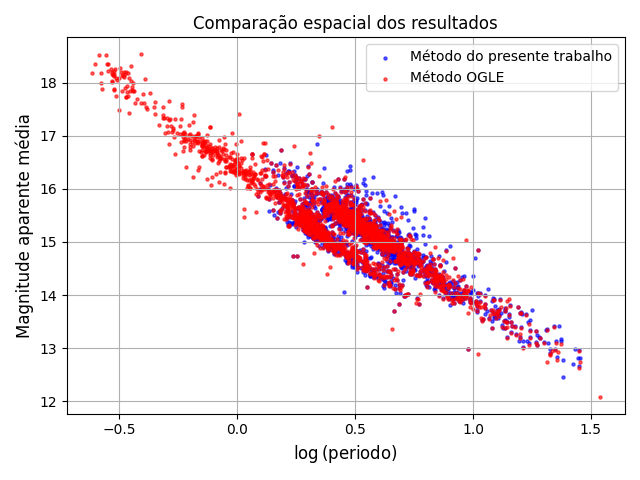}
	\caption{Lei de Leavitt para a Grande Nuvem de Magalhães com os resultados obtidos pela colaboração OGLE-IV e por esse trabalho para os dados localizados a uma distância euclidiana máxima de $10^{-1}$ entre seus pontos.  Criado por um dos autores deste trabalho (KMC).}
	\label{fig:distancia-maxima}
\end{figure}

Com o propósito de analisar e identificar padrões e relações significativas entre os resultados obtidos no estudo do OGLE-IV e os resultados deste trabalho, elaboramos o gráfico apresentado na Figura \ref{fig:distancia-maxima}. Esse gráfico tem como finalidade comparar os dois conjuntos de dados. Para realizar essa comparação, aplicamos uma restrição baseada na distância máxima.

Em nosso método, cada ponto no primeiro conjunto de dados (azuis) é comparado com todos os pontos no segundo conjunto (vermelhos) usando a norma Euclidiana para calcular a distância entre eles. Se a distância entre um ponto azul e qualquer ponto vermelho for menor ou igual à distância máxima predefinida, consideramos esse par de pontos e o representamos no gráfico. Essa abordagem nos permite identificar pontos com características semelhantes em ambos os conjuntos de dados, facilitando a análise das semelhanças entre eles.

\begin{figure}[H]
	\centering
	\includegraphics[scale=0.5]{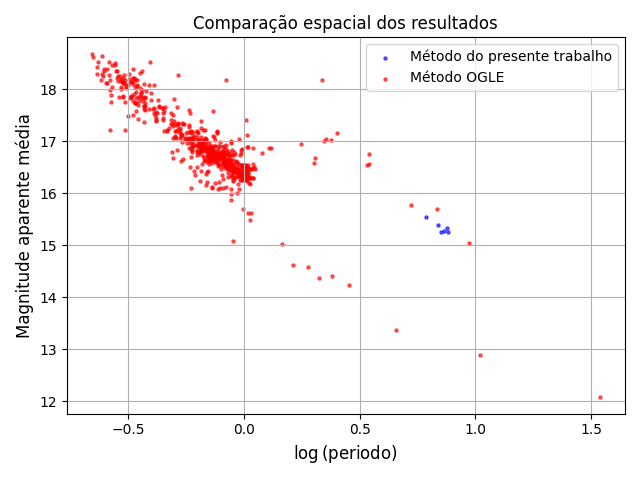}
	\caption{Lei de Leavitt para a Grande Nuvem de Magalhães com os resultados obtidos pela colaboração OGLE-IV e por esse trabalho para os dados não localizados a uma distância euclidiana maxima de $10^{-1}$ entre seus pontos.  Criado por um dos autores deste trabalho (KMC).}
	\label{fig:distancia-minima}
\end{figure}

Da mesma forma que foi feito na Figura \ref{fig:distancia-maxima}, elaboramos o gráfico apresentado na Figura \ref{fig:distancia-minima}, com uma abordagem análoga. Neste caso, os pontos que estão a uma distância euclidiana menor do que $10^{-1}$ foram removidos do conjunto de dados.

\begin{figure}[H]
	\centering
	\includegraphics[scale=0.5]{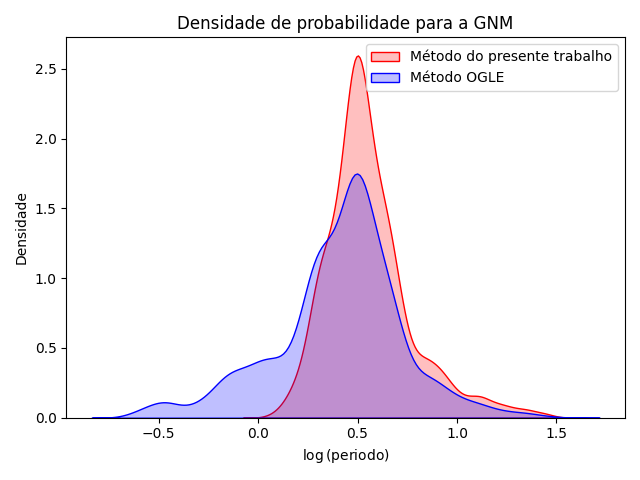}
	\caption{Densidade acumulada de logarítmo do período para os resultados do projeto OGLE-IV e desse trabalho.  Criado por um dos autores deste trabalho (KMC).}
	\label{fig:densidade-per}
\end{figure}

Com o objetivo de representar visualmente a distribuição de probabilidades dos períodos das Variáveis Cefeidas na Grande Nuvem de Magalhães, utilizamos uma técnica chamada Estimativa de Densidade por Kernel. Essa técnica consiste em suavizar os dados, aplicando funções matemáticas conhecidas como "kernels"\:a cada ponto dos conjuntos de dados, que, no caso, são os registros dos períodos em forma de logaritmo. Esses "kernels"\:são essencialmente pequenas curvas que se ajustam a cada ponto de dado, e então, são somados para criar as curvas de densidade suavizadas, que ilustram a distribuição de probabilidade dos períodos das Variáveis Cefeidas na Grande Nuvem de Magalhães. Essa abordagem estatística é amplamente reconhecida na literatura científica, conforme sugerido por Rosenblatt em seu trabalho de 1956 \cite{rosenblatt1956remarks}. A Figura \ref{fig:densidade-per} mostra o resultado desse processo.

\begin{figure}[H]
	\centering
	\includegraphics[scale=0.5]{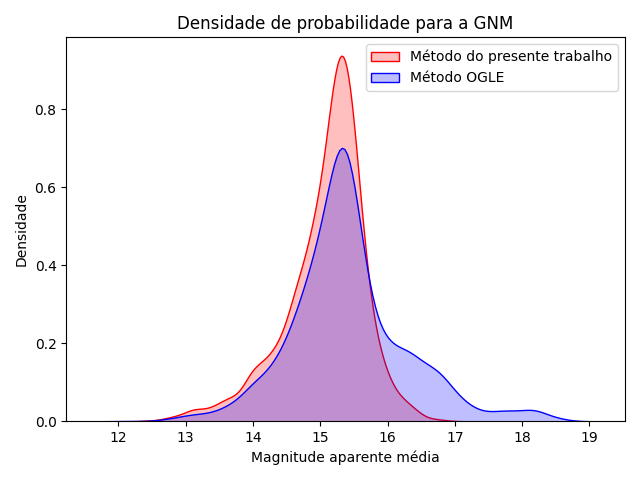}
	\caption{Densidade acumulada de magnitude aparente média para os resultados do projeto OGLE-IV e desse trabalho.  Criado por um dos autores deste trabalho (KMC).}
	\label{fig:densidade-mag}
\end{figure}

Também calculamos a densidade acumulada da magnitude aparente média das estrelas para os resultados do projeto OGLE, bem como para os resultados deste estudo. Isso é representado na Figura \ref{fig:densidade-mag}.

A análise comparativa entre os resultados deste estudo e os dados obtidos pela colaboração OGLE revelou uma notável semelhança. É relevante mencionar que esta comparação foi realizada principalmente de forma visual, sem uma análise detalhada através de testes estatísticos. Observou-se uma congruência impressionante nos padrões de distribuição de densidade de probabilidade e nas comparações diretas entre os conjuntos de dados, tudo avaliado visualmente. Este achado é particularmente significativo, pois valida a confiabilidade dos resultados apresentados neste trabalho, destacando a qualidade e integridade das informações coletadas.

\section{Determinando a Distância da Grande Nuvem de Magalhães à Via Láctea}
\label{sec:determinando.distancia.grande.nuvem.magalhaes.via.lactea}

No ano 137 A.E.C, o astrônomo Hiparco desenvolveu uma escala de classificação para as estrelas, com base em sua luminosidade aparente vista a olho nu. Ele organizou as estrelas em uma escala que variava de 1 (as mais brilhantes) a 6 (as menos brilhantes). Essa abordagem proporcionou uma maneira sistemática de descrever e catalogar as estrelas com base na intensidade de sua luminosidade percebida. Essa classificação desempenhou um papel fundamental na astronomia da época e serviu como alicerce para estudos subsequentes.

Muitos séculos mais tarde, em 1850, o astrônomo Pogson fez uma contribuição notável para a astronomia ao introduzir a escala de magnitude. Essa escala representou um marco importante na história da astronomia, estabelecendo uma relação fundamental entre a magnitude aparente, a magnitude absoluta e a distância de um objeto em parsecs, conforme mencionado em \cite{unsold2005newcosmos}. A escala de magnitude é definida por

\begin{equation}
m - M = 5 \left( \log \left(\frac{d_l}{1 parsec} \right) - 1 \right).
\label{eq:mu.r}
\end{equation}

\par Na Equação \ref{eq:mu.r} a magnitude aparente ($m$) é uma medida relativa da luminosidade de um objeto quando observado da Terra e quanto menor o valor da magnitude aparente, mais brilhante o objeto parece ser a olho nu. Por outro lado, a magnitude absoluta ($M$) é uma medida padronizada da luminosidade intrínseca do objeto, que não depende da sua distância. A equação também inclui a distância de luminosidade ($d_l$), medida em parsecs, permitindo aos astrônomos comparar objetos com base em sua verdadeira luminosidade.

Pogson reconheceu a importância da relação entre magnitude aparente, magnitude absoluta e distância como um elemento fundamental na compreensão do comportamento da luz estelar à medida que viaja pelo espaço. Essa relação se tornou uma ferramenta essencial para os astrônomos, permitindo não apenas a determinação de distâncias, mas também a classificação de diferentes tipos de estrelas com base em sua luminosidade intrínseca, conforme discutido em \cite{carroll2016modernastrophysics}. A grandeza $m - M$ é denominada módulo de distância e é representada pelo letra grega $\mu$, onde
\begin{equation}
    \mu = m - M.
    \label{eq:mu}
\end{equation}

Se tomamos um módulo de distância $\mu_X$ para um objeto utilizando um método $X$ e um módulo de distância $\mu_Y$ para o mesmo objeto tomando um método $Y$ então

\begin{equation}
    \mu_X = m_X - M_X = -5 \left(1 - \log  \left(\frac{{d_l}_X}{1 parsec} \right) \right)
\end{equation}

e 

\begin{equation}
    \mu_Y = m_Y - M_Y = -5 \left(1 - \log  \left(\frac{{d_l}_Y}{1 parsec} \right) \right),
\end{equation}

definimos também a grandeza

\begin{equation}
    \delta \mu = \mu_X - \mu_Y
\end{equation}
e, portanto,

\begin{equation}
    \delta \mu = m_X - m_Y + (M_Y - M_X).
    \label{eq:muxy}
\end{equation}

Visto que, por definição, $M$ é o brilho intríseco da estrela então $M_Y = M_X$ modificando a Equação \ref{eq:muxy} para
\begin{equation}
    \delta \mu = m_X - m_Y.
\end{equation}

Além disso, $\delta\mu$ pode também ser representado por

\begin{equation}
    {d_l}_Y = {d_l}_X 10^{\displaystyle\frac{\delta\mu}{5}}
\end{equation}

Para as estrelas Variáveis Cefeidas, o brilho varia ao longo do tempo, portanto assumiremos $m_X = \Bar{m}_{X}$ e $m_Y = \Bar{m}_{Y}$, então

\begin{equation}
    \delta\mu = \Bar{m}_{X} - \Bar{m}_{Y}.
\end{equation}

Se o método $Y$ for um método que leva em consideração a Lei de Leavitt, então

\begin{equation}
    {d_l}_\text{Leavitt} = {d_l}_X 10^{\displaystyle\frac{\delta\mu}{5}},
    \label{eq:distancia_leavitt}
\end{equation}

onde 

\begin{equation}
    \delta\mu = \Bar{m}_{X} - (A \log P + B).
\end{equation}

\par No presente trabalho, a grandeza ${m}_{X}$ e a grandeza $t$ serão as grandezas observacionais obtidas a partir do catálogo OGLE-IV, as grandezas $P$, $A$ e $B$ são grandezas cálculadas a partir da métodologia apresentada nesse trabalho, e a grandeza ${d_l}_X$ será uma distância de calibração obtida a partir de um método independente: o método das binárias eclipsantes \cite{pietrzynski2019distance}.

O método das binárias eclipsantes é uma técnica astronômica para o cálculo de distâncias a galáxias e sistemas estelares distantes. Ele se baseia na observação de sistemas estelares binários, nos quais duas estrelas orbitam uma em torno da outra. Quando uma dessas estrelas passa na frente da outra em relação à Terra, ocorre um eclipse, causando uma diminuição temporária no brilho total do sistema. Ao estudar esses eclipses e analisar variações periódicas no brilho, os astrônomos podem determinar as massas, tamanhos e luminosidades das estrelas envolvidas. Combinando essas informações com modelos físicos, eles podem calcular as distâncias até esses sistemas com grande precisão, o trabalho de Pietrzynsky utiliza esse método para o cálculo da distância à Grande Nuvem de Magalhães obtendo uma distância com um erro de $1,27\%$ \cite{pietrzynski2019distance}.

O método apresentado se fundamenta na definição da grandeza \(\delta \mu\). Esta grandeza relaciona a magnitude média com a Lei de Leavitt e pode representar a variação na luminosidade intrínseca das estrelas observadas. Ao utilizar a distância de calibração $d_l = (49.59 \pm 0.09 \, \text{(\textit{estatístico})} \pm 0.54 \, \text{(\textit{sistemático})}) \, \text{k}\,\text{pc}$ obtida pelo método das binárias eclipsantes \cite{pietrzynski2019distance}, podemos aplicar o método desenvolvido para calcular com precisão as distâncias até a Grande Nuvem de Magalhães, proporcionando uma abordagem precisa e confiável para determinar distâncias astronômicas, no presente trabalho foram obtidos dois valores para a distância à Grande Nuvem de Magalhães: $ 50,57 \pm 0,91\;\text{k}\,\text{pc}$ e $51,80 \pm 0,85\;\text{k}\,\text{pc}$.

\subsection{Distância até a Grande Nuvem de Magalhães}
\label{sec:distance}
\begin{figure}[H]
    \centering
    \begin{minipage}{0.5\textwidth}
        \centering
        \includegraphics[scale=0.4]{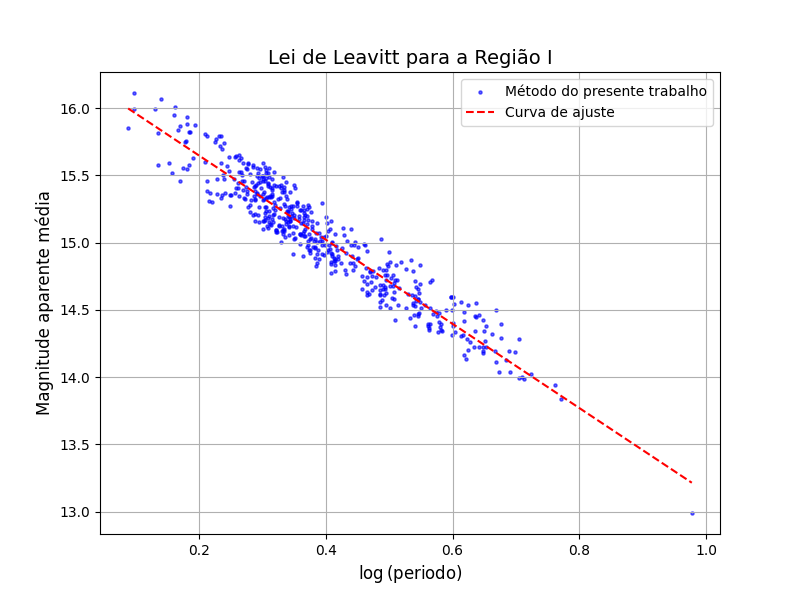}
    \end{minipage}%
    \begin{minipage}{0.5\textwidth}
        \centering
        \includegraphics[scale=0.4]{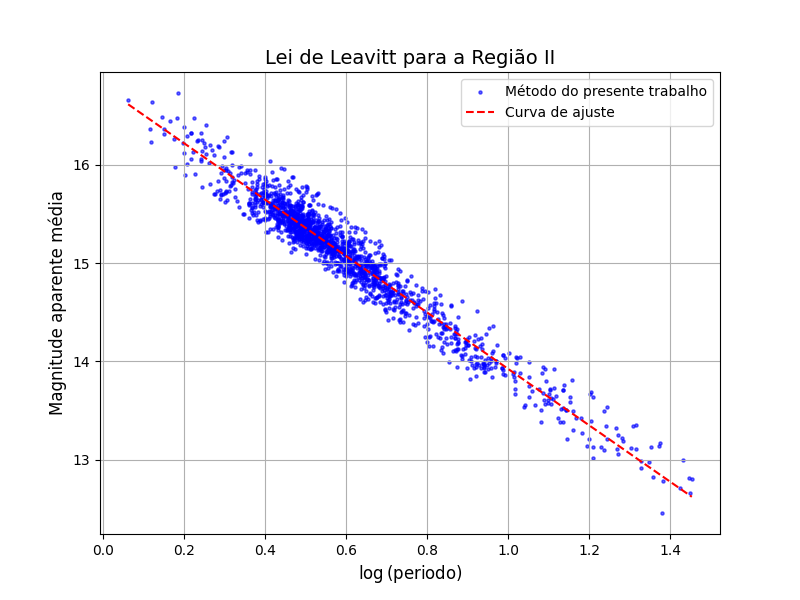}
    \end{minipage}
    \caption{Lei de Leavitt para a Região I e para a Região II com os resultados obtidos por esse trabalho.  Criado por um dos autores deste trabalho (KMC).}
    \label{fig:leavitt-regiao}
\end{figure}

Procuraremos agora determinar a distântcia da Grande Nuvem de Magalhães usando as técnicas que desenvolvemos anteriormente. Compararemos em seguida os resultados com os obtidos pela colaboração OGLE.
Com base na Figura \ref{fig:regioes}, a porção inferior será designada como Região I e a parte superior será denominada Região II, onde as Leis de Leavitt correspondentes são ilustradas na Figura \ref{fig:leavitt-regiao}.

As Leis de Leavitt obtidas para a Região I (Equação (\ref{eq:leavitt-I})) e para a Região II (Equação (\ref{eq:leavitt-II})) foram 
\begin{equation}
    m = (-3,13 \pm 0,04)\cdot \log P + (16,274 \pm 0,018)
    \label{eq:leavitt-I}
\end{equation} 
e
\begin{equation}
    m = (-2,871 \pm 0,017)\cdot \log P + (16,795 \pm 0,011).
    \label{eq:leavitt-II}
\end{equation}

\begin{figure}[H]
    \centering
    \begin{minipage}{0.5\textwidth}
        \centering
        \includegraphics[scale=0.4]{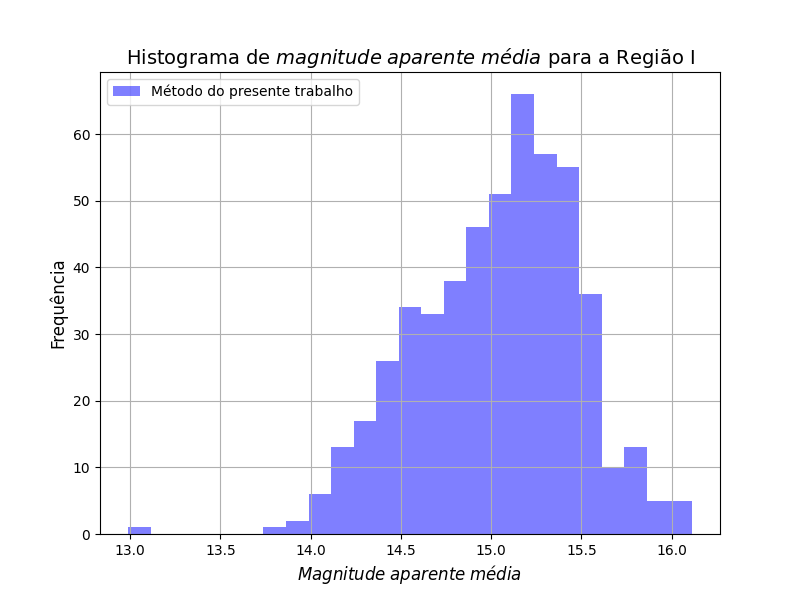}
    \end{minipage}%
    \begin{minipage}{0.5\textwidth}
        \centering
        \includegraphics[scale=0.4]{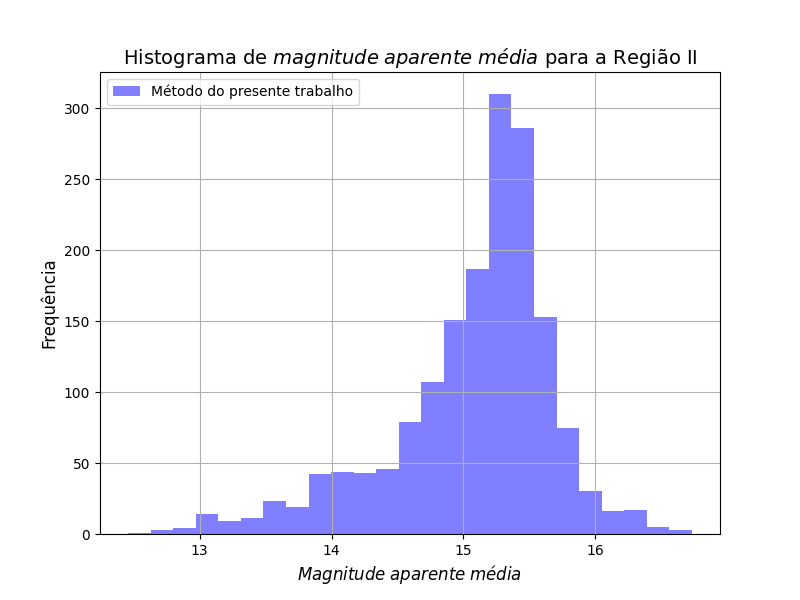}
    \end{minipage}
    \caption{Histograma de magnitude aparente média para a Região I e para a Região II com base nos resultados apresentados neste trabalho.  Criado por um dos autores deste trabalho (KMC).}
    \label{fig:dist-mag}
\end{figure}

Os histogramas de magnitude aparente média para a Região I e para a Região II estão representadas na Figura \ref{fig:dist-mag}.

\begin{figure}[H]
    \centering
    \begin{minipage}{0.5\textwidth}
        \centering
        \includegraphics[scale=0.4]{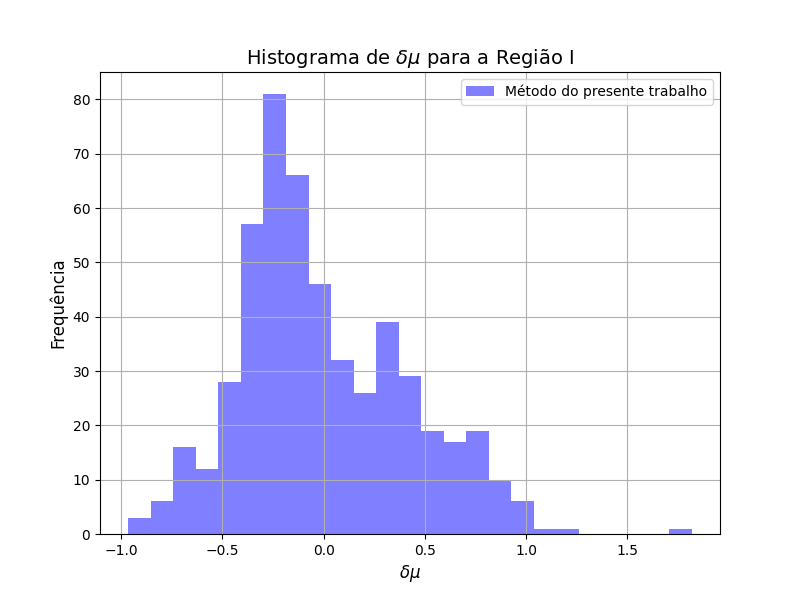}
    \end{minipage}%
    \begin{minipage}{0.5\textwidth}
        \centering
        \includegraphics[scale=0.4]{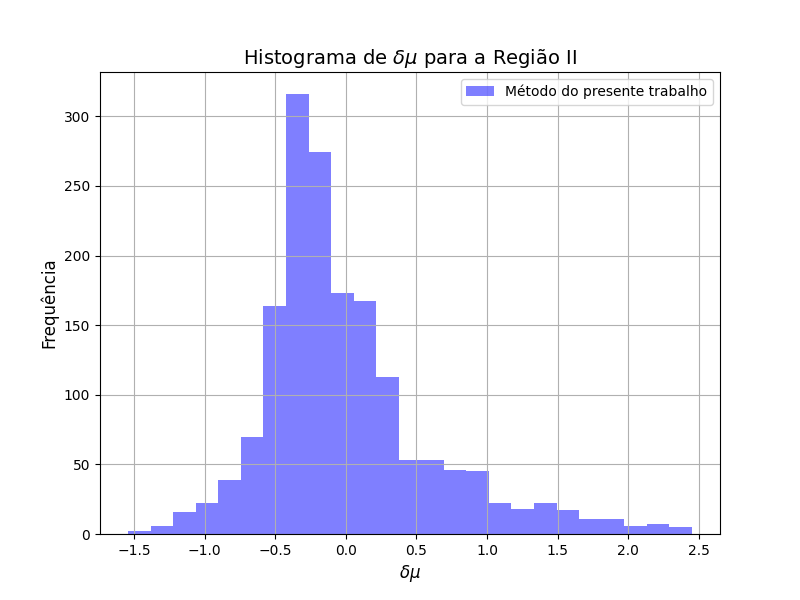}
    \end{minipage}
    \caption{Histograma de $\delta\mu$ para a Região I e para a Região II com base nos resultados apresentados neste trabalho.  Criado por um dos autores deste trabalho (KMC).}
    \label{fig:distribuicao.delta}
\end{figure}

Com base nos resultados obtidos na Seção \ref{sec:analise} e na Seção \ref{sec:determinando.distancia.grande.nuvem.magalhaes.via.lactea}, o cálculo de $\delta \mu$ para cada estrela da Região I e da Região II resulta nos histogramas representados na Figura \ref{fig:distribuicao.delta}.

\begin{figure}[H]
    \centering
    \begin{minipage}{0.5\textwidth}
        \centering
        \includegraphics[scale=0.4]{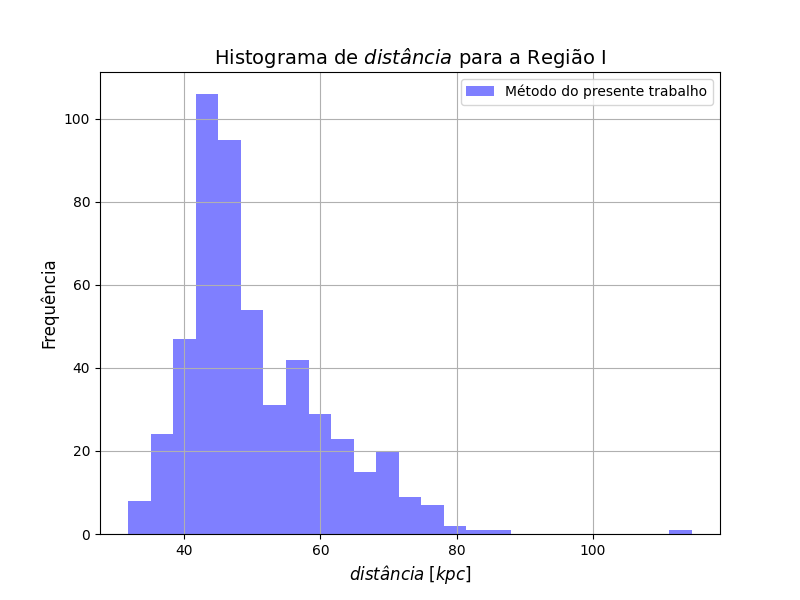}
    \end{minipage}%
    \begin{minipage}{0.5\textwidth}
        \centering
        \includegraphics[scale=0.4]{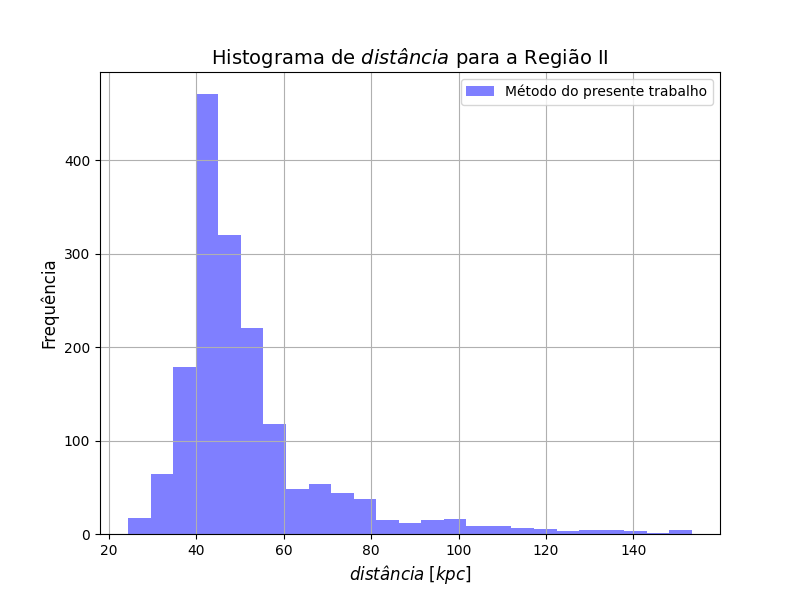}
    \end{minipage}
    \caption{Histograma de distância para a Região I e para a Região II com base nos resultados apresentados neste trabalho.  Criado por um dos autores deste trabalho (KMC).}
    \label{fig:dist-dist}
\end{figure}

Usando a Equação \ref{eq:distancia_leavitt} e considerando a distância $d_l = (49.59 \pm 0.09 \, \text{(\textit{estatístico})} \pm 0.54 \, \text{(\textit{sistemático})}) \, \text{k}\,\text{pc}$ do método das binárias eclipsantes - conforme discutido no início dessa seção - como a distância de referência, obtém-se a distribuição de distâncias para a Região I e a Região II, conforme ilustrado na Figura \ref{fig:dist-dist}.

Com base nos resultados dos gráficos apresentados na Figura \ref{fig:distribuicao.delta} e na Figura \ref{fig:dist-dist}, é possível realizar a análise da relação entre a distância das estrelas da Região I e da Região II e a grandeza $\delta\mu$.

\begin{figure}[H]
    \centering
    \begin{minipage}{0.5\textwidth}
        \centering
        \includegraphics[scale=0.4]{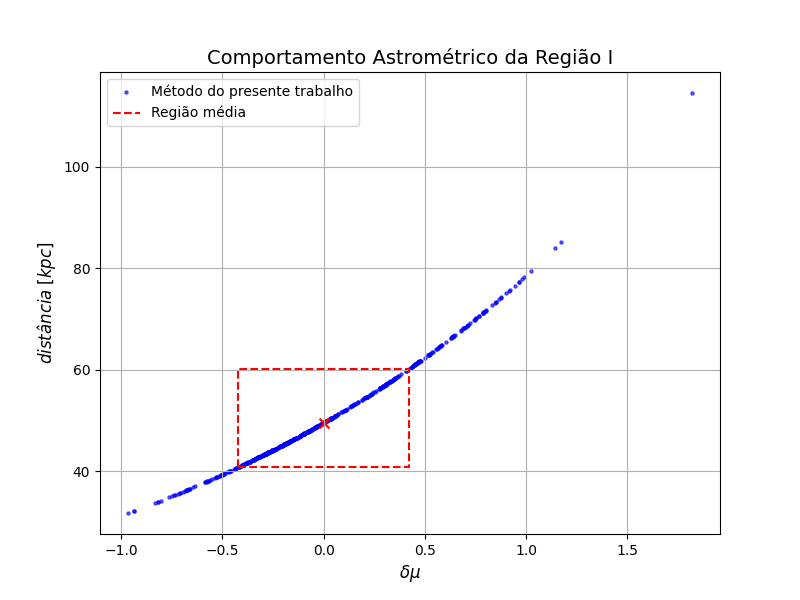}
    \end{minipage}%
    \begin{minipage}{0.5\textwidth}
        \centering
        \includegraphics[scale=0.4]{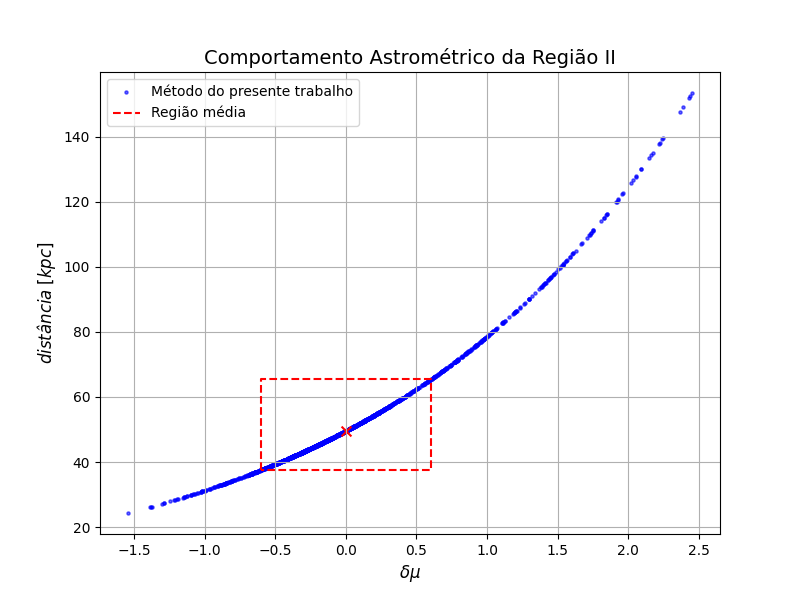}
    \end{minipage}
    \caption{Relação entre a distância de luminosidade e a grandeza $\delta\mu$ para as estrelas da Região I e da Região II, com base nos resultados apresentados neste trabalho.  Criado por um dos autores deste trabalho (KMC).}
    \label{fig:relacao.mu.distance}
\end{figure}

A Tabela \ref{TabelaResultados} apresenta os resultados obtidos tanto para a Região I quanto para a Região II. 

\begin{table}[H]
\centering
\caption{Resultados para as Regiões I e II, com um intervalo de confiança de 95\%.}
\label{TabelaResultados}
\begin{tabular}{|c|c|c|c|c|c|c|}
\hline
 & Distância Média & Distância Mínima & Distância Máxima & $\delta\mu$ Médio & Quantidade de Estrelas \\
\hline
Região I & $ 50,57 \pm 0,91\;\text{k}\,\text{pc}$ & $31,79 \; \text{k}\,\text{pc}$  &  $114,54 \;\text{k}\,\text{pc}$ & $0,00 \pm 0,036$& $515$ \\
\hline
Região II & $51,80 \pm 0,85\;\text{k}\,\text{pc}$ &$24,37 \; \text{k}\,\text{pc}$ &$153,38 \; \text{k}\,\text{pc}$  & $0,00 \pm 0,029$ & $1678$ \\
\hline
\end{tabular}
\end{table}

A colaboração OGLE-IV também possui trabalhos para o cálculo da distância à Grande Nuvem de Magalhães, porém o OGLE-IV adota uma métodologia diferente da abordada nesse trabalho, eles utilizam como distância de calibração a distância de $d_{l} = 49,97 \pm 0,19 (\mathrm{estatistico}) \pm 1,11 (\mathrm{sistematico}) \; \text{k}\,\text{pc}$ obtída pelo método das binárias eclipsantes do trabalho de Pietrzynski, de 2013 \cite{pietrzynski2013}.\\

\par 

Além disso, a definição do $\delta\mu$ é diferente:
\begin{equation}
    \delta\mu = W_{I, V-I} - ( a \log (P) + b),
\end{equation}
onde
\begin{equation}
    W_{I, V-I} = m_I - 1,44 (m_V - m_I)
\end{equation}
é a denominada magnitude de Wesenheit: a magnitude de Wesenheit é uma magnitude corrigida, calculada a partir de magnitudes aparentes em diferentes bandas de cores, geralmente no infravermelho próximo, onde a extinção do meio interestelar é significativamente menor do que na luz visível \cite{ngeow2012}.

É observado que a relação entre o período e a luminosidade exibe comportamento semelhante tanto na Grande Nuvem de Magalhães quanto na Pequena Nuvem de Magalhães quando o valor de $\log (P)$ é inferior a 0,4. No entanto, é importante notar que o catálogo mantido pela colaboração OGLE registra um número significativamente maior de estrelas Variáveis Cefeidas na Pequena Nuvem de Magalhães em comparação com a Grande Nuvem de Magalhães. Assumir a mesma relação para ambas as galáxias poderia potencialmente introduzir um viés nos resultados.

Além da discrepância no número de Variáveis Cefeidas, é crucial levar em conta que a Grande Nuvem de Magalhães e a Pequena Nuvem de Magalhães são diferentes em termos de composição química e propriedades locais. Portanto, é intrigante que ambas manifestem o mesmo comportamento na faixa de períodos em questão. Por esse motivo, a colaboração OGLE-IV opta por não incluir os dados correspondentes a $\log (P) < 0,4$ ao calcular a distância para a Grande Nuvem de Magalhães \cite{jacyczyn2016ogleing}.

\section{Conclusão}

Neste estudo, determinamos a distância até a Grande Nuvem de Magalhães aplicando a Lei de Leavitt às Variáveis Cefeidas disponíveis no catálogo OGLE-IV. Para calcular a Lei de Leavitt, foi fundamental determinar o período de oscilação das Variáveis Cefeidas. Utilizamos o Periodograma Lomb-Scargle, implementado na linguagem de programação Python 3.9, para realizar essa tarefa. Após obter os períodos, conduzimos uma análise da dispersão nas curvas de luz de cada Variável Cefeida, calculando um coeficiente de dispersão e selecionando as Variáveis Cefeidas adequadas para nossa análise.

Após a classificação das estrelas, observamos que na Grande Nuvem de Magalhães existem Variáveis Cefeidas que pulsam de duas maneiras distintas e estão localizadas em duas regiões separadas do diagrama período-luminosidade. Essas áreas foram identificadas como Região I, que inclui 515 Variáveis Cefeidas, e Região II, com 1678 Variáveis Cefeidas. Nossos resultados foram comparados com o diagrama período-luminosidade calculado pela colaboração OGLE, e os resultados se mostraram consistentes com as descobertas da equipe OGLE.

Usando as Variáveis Cefeidas das Regiões I e II, derivamos a Lei de Leavitt para cada uma delas. Para a Região I, a relação encontrada foi
\[m = (-3,13 \pm 0,04)\cdot \log P + (16,274 \pm 0,018),\]
enquanto para a Região II, a Lei de Leavitt obtida foi
\[m = (-2,871 \pm 0,017)\cdot \log P + (16,795 \pm 0,011).\]

 \par Após estabelecermos a Lei de Leavitt, introduzimos a variável $\delta \mu$, que conecta a distância de luminosidade na banda I com uma distância física de calibração. Utilizamos a medida física $d_{l}$, que foi determinada como $ (49.59 \pm 0.09 \, \text{(\textit{erro estatístico})} \pm 0.54 \, \text{(\textit{erro sistemático})}) \, \text{k}\,\text{pc}$, obtida com precisão de $1.27\%$ pelo método das binárias eclipsantes \cite{pietrzynski2019distance}. Com base nesses dados, calculamos as distâncias individuais das Variáveis Cefeidas da Grande Nuvem de Magalhães até a Via Láctea. Em seguida, determinamos a distância média para ambas as Regiões I e II da Grande Nuvem de Magalhães. A distância média calculada para a Região I foi de $ (50,57 \pm 0,91)\text{k}\,\text{pc}$, enquanto para a Região II foi de $ (51,80 \pm 0,85)\text{k}\,\text{pc}$.

A colaboração OGLE utiliza dados das Variáveis Cefeidas para mapear a estrutura tridimensional das Nuvens de Magalhães. Para esse propósito, eles também determinam as distâncias individuais das Variáveis Cefeidas em sua base de dados. Contudo, adotam uma abordagem metodológica distinta em comparação com este estudo. Na obtenção da grandeza $\delta\mu$, a colaboração OGLE emprega a Magnitude de Wesenheit, uma magnitude corrigida derivada da magnitude aparente em diferentes bandas, especialmente no infravermelho próximo, onde a extinção do meio interestelar é significativamente menor. Para calcular essa magnitude, eles utilizam Variáveis Cefeidas com medidas tanto na banda I quanto na banda V. É crucial notar que nem todas as Variáveis Cefeidas no catálogo OGLE possuem medidas em ambas as bandas simultaneamente, implicando que as estrelas utilizadas pelo OGLE diferem daquelas empregadas neste estudo.

Além disso, a colaboração OGLE estabelece um critério de seleção que engloba apenas Variáveis Cefeidas com $\log P < 0.4$. Consequentemente, várias das Variáveis Cefeidas analisadas neste estudo não são incluídas na análise da equipe OGLE. Além disso, neste trabalho, o processo de cálculo do período, especialmente no que diz respeito ao coeficiente de dispersão, bem como o método de seleção das estrelas, segue uma abordagem metodológica distinta. Utilizamos o Periodograma Lomb-Scargle e implementamos um método de análise de imagens para classificar as Variáveis Cefeidas. Essas discrepâncias na metodologia e, sobretudo, nas Variáveis Cefeidas examinadas, são responsáveis pela distinção fundamental entre os dois estudos.

Esperamos que este artigo possa se tornar uma ferramenta valiosa para estudantes de diferentes níveis acadêmicos que estejam interessados em iniciar sua jornada no estudo das Variáveis Cefeidas. Ele oferece uma introdução clara aos cálculos, começando desde a determinação do período de pulsação das estrelas até o cálculo da distância, de maneira progressiva e lógica. Com esse propósito, disponibilizamos o programa utilizado para calcular o período de pulsação em um repositório no GitHub, acessível pelo link \url{https://github.com/Costa-Kevin-M/distancias_galacticas}. Desta forma, os estudantes podem seguir a metodologia apresentada neste trabalho, reproduzir nossos resultados e compreender minuciosamente cada etapa do processo.

\section{Agradecimentos}

Os autores agradecem CNPq e FAPES pelo apoio financeiro parcial. K.M.C. gostaria de agradecer, em especial, a bolsa concedida pelo CNPq que permitiu desenvolver este trabalho.

\bibliography{bibliografia}

\end{document}